\documentclass[fleqn,usenatbib]{mnras}
\usepackage{newtxtext,newtxmath}
\usepackage[T1]{fontenc}
\DeclareRobustCommand{\VAN}[3]{#2}
\let\VANthebibliography\thebibliography
\def\thebibliography{\DeclareRobustCommand{\VAN}[3]{##3}\VANthebibliography}
\usepackage{ae,aecompl}
\usepackage{graphicx}   
\usepackage{subcaption} 
\usepackage{amsmath}    
\usepackage{pdflscape} 
\usepackage{hyperref}
\usepackage[dvipsnames]{xcolor}

\title[Sun-as-a-star EUV mHz]{Signatures of photospheric convection throughout the solar atmosphere: the EVE Sun-as-a-star mHz continuum}

\author[H.~S. Hudson, \textit{et al.}]{
Hugh Hudson,$^{1,2}$\thanks{E-mail: Hugh.Hudson@glasgow.ac.uk}
\href{https://orcid.org/0000-0001-5685-1283}
{\includegraphics[scale=0.07]{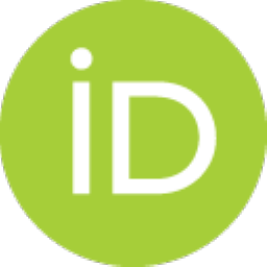}}
Anne-Marie Broomhall,$^{3}$
Lyndsay Fletcher,$^{1,4}$ 
David Graham,$^{1}$
\newauthor
Sargam M. Mulay,$^{1}$\thanks{E-mail: Sargam.Mulay@glasgow.ac.uk}
\href{https://orcid.org/0000-0002-9242-2643}
{\includegraphics[scale=0.07]{Figures/img/ORCIDiD_icon.pdf}}
Chris Osborne,$^{1}$
Kirsty Williamson,$^{1}$ and
Graham Woan$^{1}$ \\
$^{1}$SUPA School of Physics \& Astronomy, University of Glasgow, Glasgow G12 8QQ, UK\\ 
$^{2}$Space Science Laboratory, University of California, Berkeley, CA 94720 USA\\
$^{3}$University of Warwick, UK \\
$^{4}$Rosseland Centre for Solar Physics, University of Oslo, PO Box 1029 Blindern, NO-0315 Oslo, Norway
}

\date{Accepted XXX. Received YYY; in original form ZZZ}
\pubyear{WWW}

\begin{document}

\label{firstpage}
\pagerange{\pageref{firstpage}--\pageref{lastpage}}

\maketitle
\begin{abstract}
Convectively driven motions in the solar photosphere can generate broadband Doppler variability across the chromosphere, transition region and corona. 
Here we investigate this variability using ``Sun-as-a-Star'' observations from the Extreme Ultraviolet Variability Experiment (EVE) aboard the Solar Dynamics Observatory, constructing high signal-to-noise Doppler power spectra by incoherently averaging spectra from 3-hour sequences of the centroid wavelengths of emission lines that span wavelengths 35–104~nm. 
The spectra reveal a broad power-spectral continuum with two ``Harvey'' components, one of which extends to the Nyquist frequency at 50~mHz with a steep power-law tail. 
All emission lines have substantial power both above and below the 
nominal acoustic cutoff frequency at 5~mHz, but none of the power spectra show maxima corresponding to p-mode or chromospheric 3-minute oscillations.
Lines formed in the corona, as compared with those of the chromosphere/transition region, have substantially less Doppler amplitude in the Harvey spectrum associated with granulation-scale convection.
Based on the observed continuum, there is no evidence (in any of the 26 lines studied) for Kolmogorov turbulence, which predicts a flat continuum component with Doppler variance $\langle{v^2}\rangle\,\propto\,f^{-5/3}$ as a function of frequency $f$.
The total inferred RMS velocities ($>0.1$~mHz) are roughly consistent with coronal ``microturbulence'' estimates from excess line widths. 
These observations provide the first clear detection of Sun-as-a-star EUV Doppler variability above $\sim$10 mHz and demonstrate the potential of full-disk EUV spectroscopy to probe turbulent energy transport throughout a stellar atmosphere.
The predominance of the Harvey spectrum, a signature of photospheric convection, likely results from limb brightening of the Doppler variance due to horizontal flows in the upper solar atmosphere.

\end{abstract} 


\section{Introduction}\label{sec:intro}

The Extreme ultraviolet (EUV) wavelengths in a stellar spectrum provide detailed information about the structure and dynamics of its atmosphere.
These wavelengths require observations from space platforms.
For the Sun, the existing observations include both imaging and spectroscopy, but for distant stars, we have only the latter.
This article discusses Sun-as-a-star observations of the solar spectrum as derived from the Extreme Ultraviolet Variability Experiment (EVE; \cite{EVE_Woods_2012}) instrument on board the Solar Dynamics Observatory (SDO; \citealt{Pesnell12}).
This experiment records spectra (starting in April 2010) across the range 5-104~nm, almost continuously, at an initial 10~s cadence, dropping to 1-min cadence at the time of writing.
\nocite{2012SoPh..275..115W}
The short-wavelength data (5-35~nm) ended in 2014.
We have found that the EVE spectroscopy spectra have great stability, and even though the spectral binning is only 0.02~nm, the precision of the data permits Doppler measurements in the range of a few km/s per 10~s sample \citep{2011SoPh..273...69H}.

\cite{2021ApJ...916...66B} describe solar variability in the full Fourier range from below 1~$\mu$Hz, emphasizing solar-stellar comparisons, broadly distinguishing active-region contributions, a ``mid-frequency continuum,'' convective-band contributions, and the p-modes.
The convective-band contributions (such as the granulation) have a simple interpretation as described by \cite{1985ESASP.235..199H}: a given spatial scale of convective motions results in a power spectrum that is flat up to a mean frequency corresponding to the spatial scale, with a roll-over to a power spectrum (nominally $f^{-2}$) at higher frequencies. 
Subtle variations in the slope of the continuum reflect different scales as reported, for example, by \cite{2003ASPC..294..441A} and by \cite{2012MNRAS.421.3170K}.
The best data for characterizing the power spectrum of the photosphere come from the lengthy time series of total solar irradiance (TSI) and the VIRGO photometers on board SOHO.
These data typically have 1-min sampling, corresponding to a Nyquist frequency of 8.33-mHz.

Photospheric convective motions must affect higher levels in the solar atmosphere. 
These motions reflect an upward flow of energy and momentum that EUV data via EVE spectroscopy can help to track.
The photospheric flows include several contributions in the mHz frequency range: convective motions, a broadband continuum in a power spectrum \citep{1985ESASP.235..199H}; the p-mode lines, for which Sun-as-a-star data can detect the low-$\ell$ members \citep{1979Natur.282..591C, 1983Natur.305..589W}; and various motions associated with flares and CMEs \citep{2011SoPh..273...69H,2018ApJ...862...59B,2022ApJ...931...76X}.
We can describe the photospheric variability as a broad-band wave source at the layer defined by optical depth unity at 500~nm in the visible spectrum: $\tau_{500}$ = 1.
The flows there excite acoustic waves, but these cannot propagate upward due to evanescence, and instead the upwelling wave power must be described in terms of MHD modes \citep{2019NatAs...3..223M}.
Recent observations from various platforms have tracked the convective continua into the chromosphere and corona \citep{2025A&A...701A.199A,2025ApJ...982..104M,2026ApJ..1001..173Q}, and even into the heliosphere \citep{2026ApJ...999L...4H}, but generally via images rather than via the global Sun-as-a-star view that EVE gives us. 

In this article, we follow \cite{2025ApJ...987..161K} in seeking to understand the vertical changes of flow fields in the solar atmosphere, via time-series studies of spectral lines formed at different altitudes.
Because EVE makes Sun-as-a-star observations, the height contribution functions are complicated, and in this article we characterize them only by their ionization-equilibrium formation temperatures as obtained from CHIANTI \citep{2019ApJS..241...22D}, and by a crude ``opacity index'' estimation.

This article makes a first general assessment of the EVE data in the p-mode/convective variability ranges.
These data provide a novel and comprehensive view of this spectrum in a broad wavelength range (35-104~nm), covering frequencies above 0.1~mHz as dictated by the availability of 3-h data chunks, up to a Nyquist frequency of 50~mHz.
This is well above the time scales of the main convective and oscillatory modes of the photosphere.
Note that most of the astronomical (exoplanet) interest is at power-spectral frequencies below 10$\mu$Hz (one day), where active-region development and 
stellar rotation dominate the variability.
Here we mainly study the January 2011 data based on daily 3-h chunks, typically using a large
fraction of the possible 365 chunks depending upon data quality.
Some information on lower frequencies is available, but will not be discussed in this article.
In most of the lines surveyed, the solar-origin Doppler spectral power density at 50~mHz dominates the white noise at high frequencies expected from sensitivity limits, and the power spectra universally exhibit a broad continuum driven ultimately by photospheric motions.
To place the EUV power spectra in this context, we first review Sun-as-a-star data from the ``ground truth'' SDO/HMI observations (Section~\ref{sec:hmi}).

The main theme of this paper is the morphology of the power spectra for the available lines.
Most of these (the ``metals'') form only in the transition region and corona, but we also have five lines of H and He, which we deal with in Section~\ref{sec:chrom}. 
In all of the lines we see substantial power below 5~mHz, consistent with the operation of the ``magnetic portal'' for acoustic waves as described by \cite{2007Sci...317.1192T}.
These Sun-as-a-star observations confirm that the portal behavior extends throughout the global chromosphere.

\section{Observations}\label{sec:obs}

\subsection{HMI ``ground truth''}\label{sec:hmi}

The Helioseismic and Magnetic Imager (HMI) instrument \citep{2012SoPh..275..207S} on SDO conveniently provides the best database characterizing the motions in the photosphere that likely drive the EUV variability in the upper atmosphere (chromosphere/transition region/low corona) that EVE observes.
We adopt HMI's Ic pseudo-continuum as a proxy for the true continuum, and make the shortcut of using image-sum data from the FITS file headers.
These have a 45-s cadence and result from the full Stokes measurements from the many polarization and wavelength spectroheliograms during each time interval.
The 45-s sampling of HMI gives a Nyquist frequency of 11.1~mHz.
The standard processing of the HMI data also gives a line-of-sight velocity product $V_{LOS}$, which is our main comparison.

\begin{figure}
\centering
    \includegraphics[width=\linewidth]{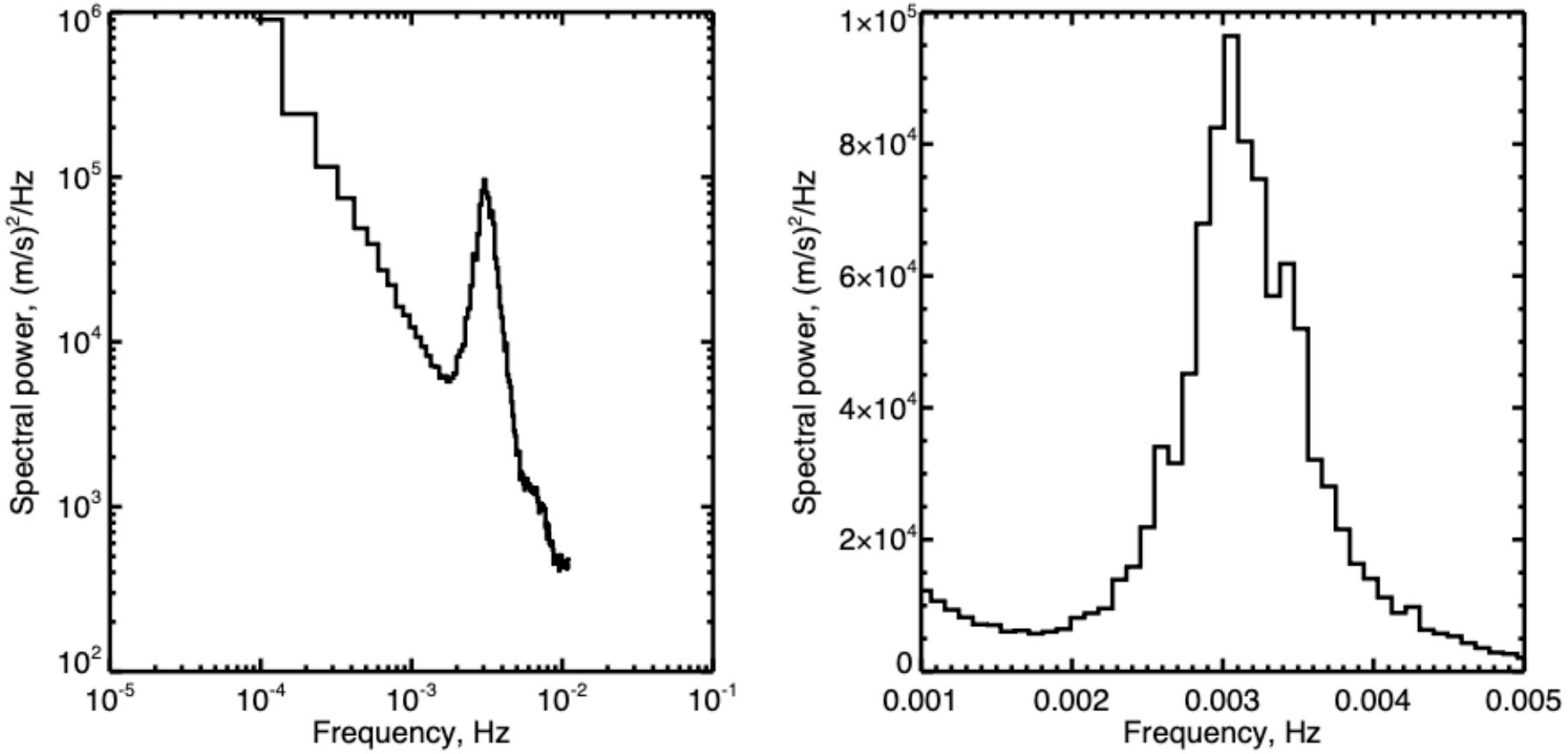}
    \includegraphics[width=\linewidth]{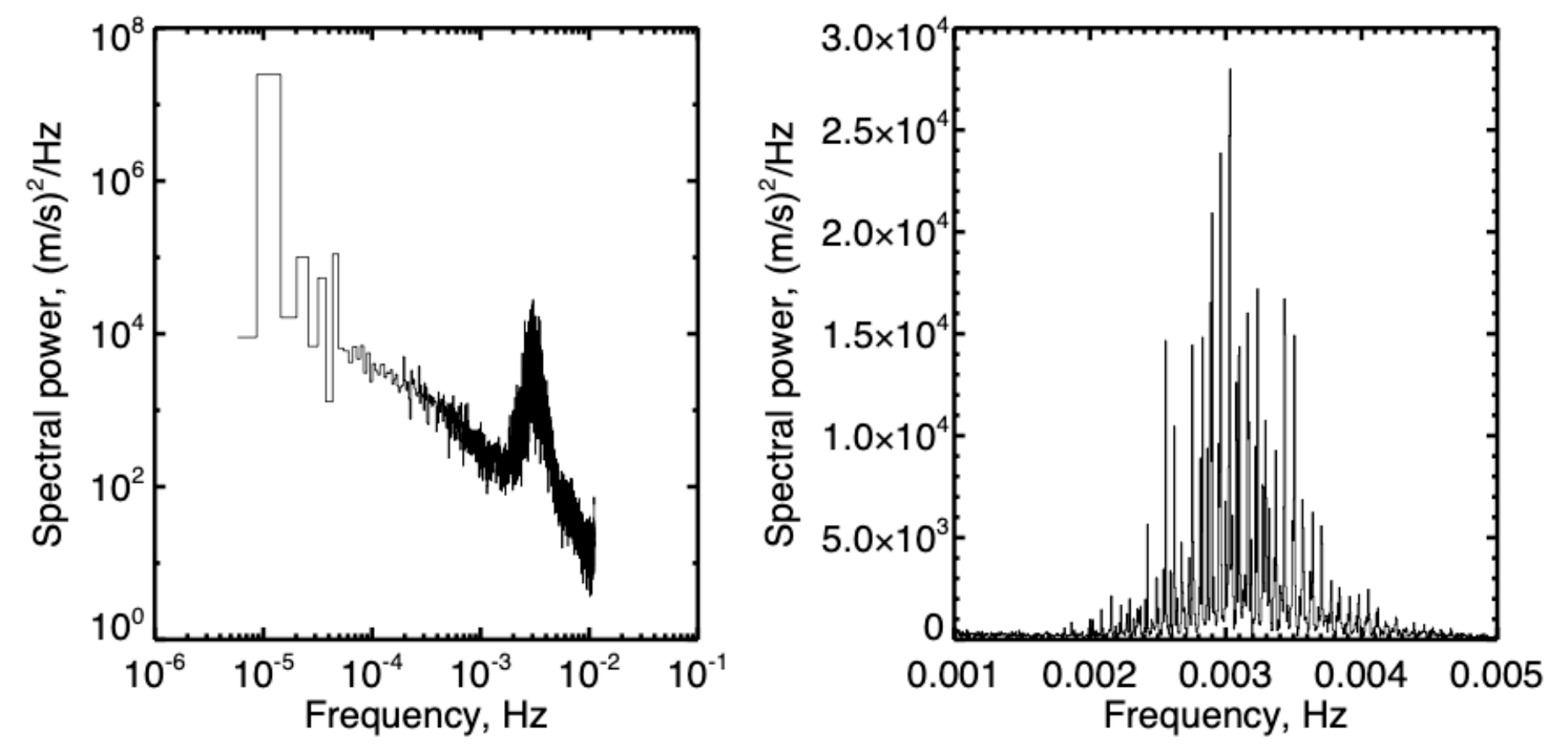}
\caption{\textit{HMI $V_{LOS}$} incoherent-sum Doppler spectra: top pair, for the 232 valid three-hour data chunks available in January 2011; bottom pair, for the 8 valid two-day data chunks in January 2011.
Note that the longer timeseries, with $f_{ny} = 6\,\mu$Hz, resolves individual peaks in the p-mode spectrum.}
\label{fig:hmi_fft}
\end{figure}

We note that a convenient software package {\sc SolAster}\footnote{\url{https://tamarervin.github.io/SolAster/}} of \cite{2022ascl.soft07009E} 
can incorporate more corrections to the HMI data.
These include not just Sun-to-SDO distance and relative velocity, which our FITS-header analysis handles, but also subtleties such as limb-darkening and solar (differential) rotation.
These refinements are important on the longer timescales relevant to exoplanet searches, but not so relevant for the mHz frequency range.
Figure~\ref{fig:hmi_fft} shows power spectra for line-of-sight velocity $V_{LOS}$, obtained from all clean HMI 3-hour and 2-day data intervals (``chunks'') separately Fourier-transformed and co-added incoherently.
We used IDL's {\sc fft\_powerspectrum.pro}, with no window function, on the residuals relative to the chunk median value.
We use the {\sc jpl\_horizons} ephemeris\footnote{\url{https://ssd.jpl.nasa.gov/orbits.html}} for Sun/SDO relative position and velocity corrections, mainly diurnal owing to SDO's geostationary orbit. 

We calculated the HMI power spectra in Figure~\ref{fig:hmi_fft} by a shortcut approach, using $V_{LOS}$ image sums unweighted by limb darkening. 
In the EUV, the weighting may be different;
photospheric motions in the low-order p-modes are mainly vertical, and therefore strongly limb-darkened.
Convective motions could have a mainly horizontal contribution and therefore could show limb brightening in image data.
The behaviour of the network component, including faculae, may also differ across the frequency spectrum \citep{2012MNRAS.421.3170K}.

\subsection{EVE data}\label{sec:eve}

\subsubsection{General properties}\label{sec:general}

\begin{figure*}
\centering
    \includegraphics[width=0.8\linewidth]{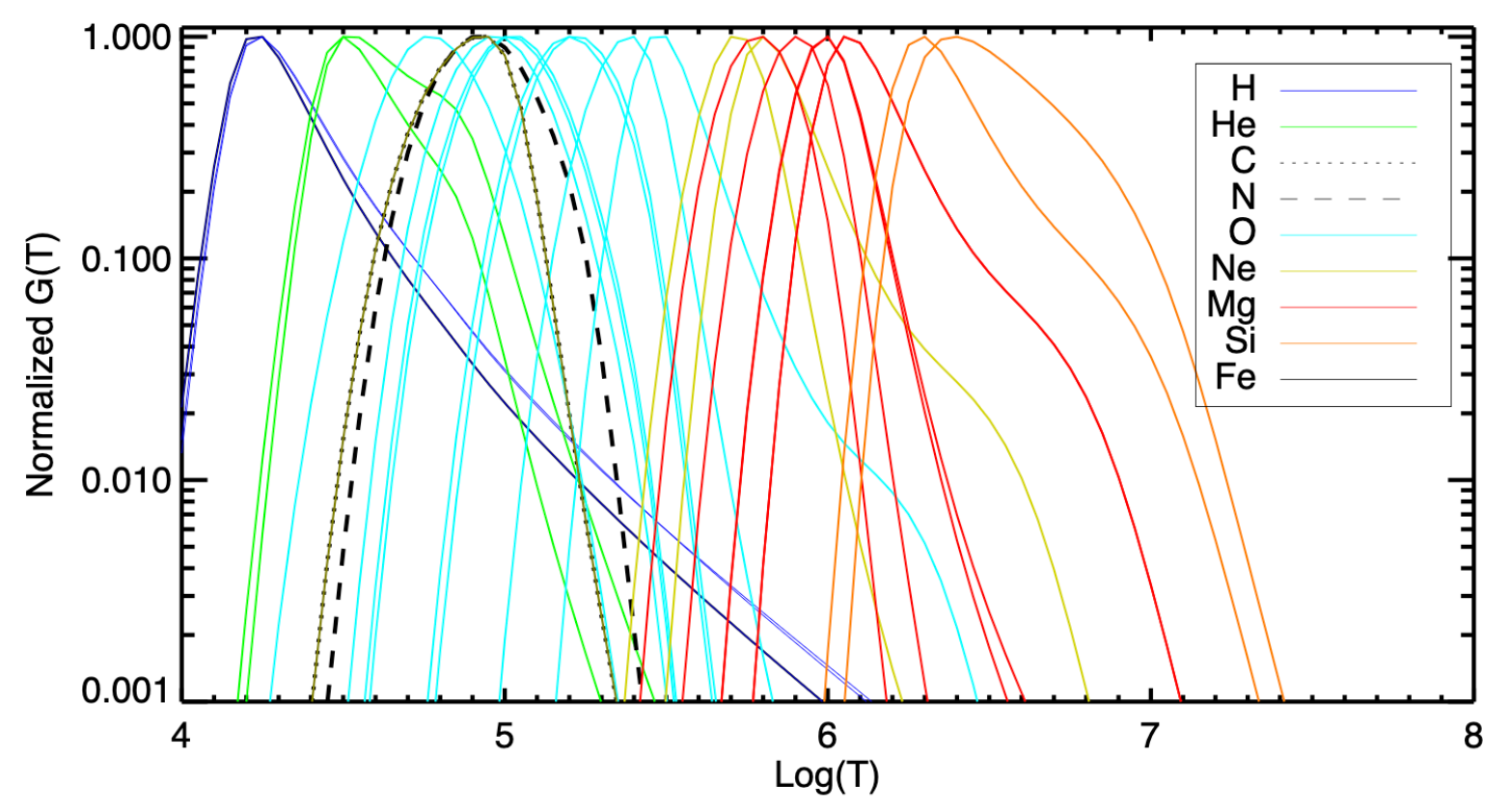}
\caption{\textit{The normalised contribution function G(T) functions for each of the lines in Table~\ref{tab:linelist}, assuming zero optical depth.
Please refer to the Table for line details.
The very prominent transition-region lines C~{\sc iii} (97.7~nm) and 
N~{\sc iii} (99.1~nm) are shown as dashed and dotted lines, respectively.
Note also the wide range of temperature sampling in the O lines (pale blue).
Generally the ionization states of the selected lines range from {\sc i} for the hydrogen Lyman lines at the left, to {\sc xii, xvi} for the Si and Fe lines at the right.}
}
\label{fig:master_goft}
\end{figure*}

In this article, we study EVE's Multiple EUV Grating Spectrograph B (MEGS-B) timeseries data, covering $\sim$33--104~nm.
Figure~\ref{fig:master_goft} shows the G(T) functions for the chosen line list (described further below), as obtained from the Chianti database\footnote{\url{https://www.chiantidatabase.org}}, illustrating how well EVE lines cover the temperature domain.

\begin{figure}
\centering
    \includegraphics[width=\linewidth]{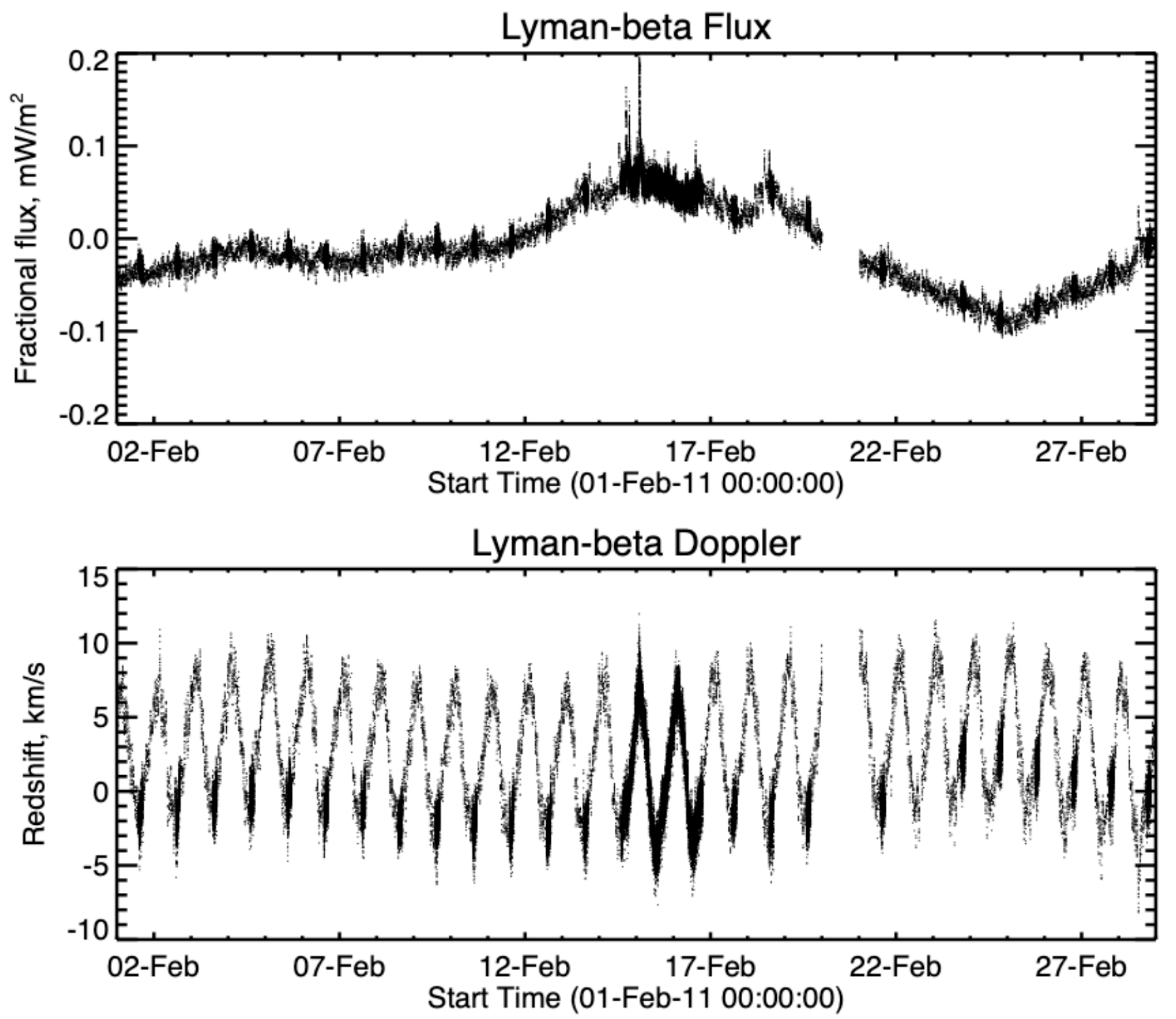}
\caption{\textit{
Lyman-$\beta$ flux (upper) and apparent wavelength (lower), expressed as Doppler redshift residuals against median values, for each 10~s EVE spectrum in February 2011.
The spectra show solar variability on active-region timescales and the X-class flare SOL2011-02-15. 
The prominent diurnal Doppler variation (about 3~km/s) results from the SDO orbital motion at geosynchronous orbit, which we correct for in further analysis.}
}
\label{fig:feb_timeseries}
\end{figure}

To provide an overview of the EVE timeseries data, we display the data for February, 2011 in
Figure~\ref{fig:feb_timeseries} as an example.
This shows the fitted fluxes and wavelengths for each 10-s sample as derived from Gaussian fits to the Lyman-$\beta$ line at 102.5~nm.
The data timeseries consists of regularly programmed 3-h intervals each day, yielding line flux and centroid estimates that we translate into Doppler shifts as residuals relative to the median values.
The roughly sinusoidal variation of line flux reflects the satellite orbital motion projected on the line of sight to the Sun, and the lower panel of Fig.~\ref{fig:feb_timeseries} gives this in Doppler redshift units of km/s.
This month saw the first X-class flare SOL2011-02-15T01:56 (X2.2), caught by EVE with a special continuous timeseries.
We see the flares in both flux and Doppler residuals \citep[cf.][]{2011SoPh..273...69H,2018ApJ...862...59B}, along with other more minor 
activity. 
The flare Doppler variations basically track mass motions during coronal evaporation, which the 10-s time sampling can follow in major events.
See \cite{2026SoPh..301...63W} for a recent discussion.

The analysis we report in this paper depends upon being able to interpolate relative wavelengths with high precision, but not necessarily high accuracy.
For the latter we would need to rely upon the absolute wavelength calibrations, and Figure~\ref{fig:feb_timeseries} already reveals one major systematic issue (the SDO diurnal motions).
In addition to this we could also note the weak wavelength dependence on source position \citep{2016SoPh..291.1665C,2026SoPh..301...63W} as well as routine secular drifts and any adjustments of the wavelength calibration.

As described below, our analysis uses three-hour data chunks and uses differential wavelengths only, via using timeseries of line centroids relative to the median value for each interval.
This means that wavelength calibration changes over time scales longer than three hours basically cancel out.
The SDO orbital motion is at a roughly diurnal (24-hour) period, and not directly relevant to the power spectrum of a single data chunk; we have nevertheless made a precise correction for the Sun-SDO distance variation  based on JPL Horizons ephemeris data.
This directly removes any unwanted variations at multiples of 11.5~$\mu$Hz from harmonics of the basic 1/day frequency.

\subsubsection{Incoherent-sum analysis}

\begin{figure}
\centering
    \includegraphics[width=\linewidth]{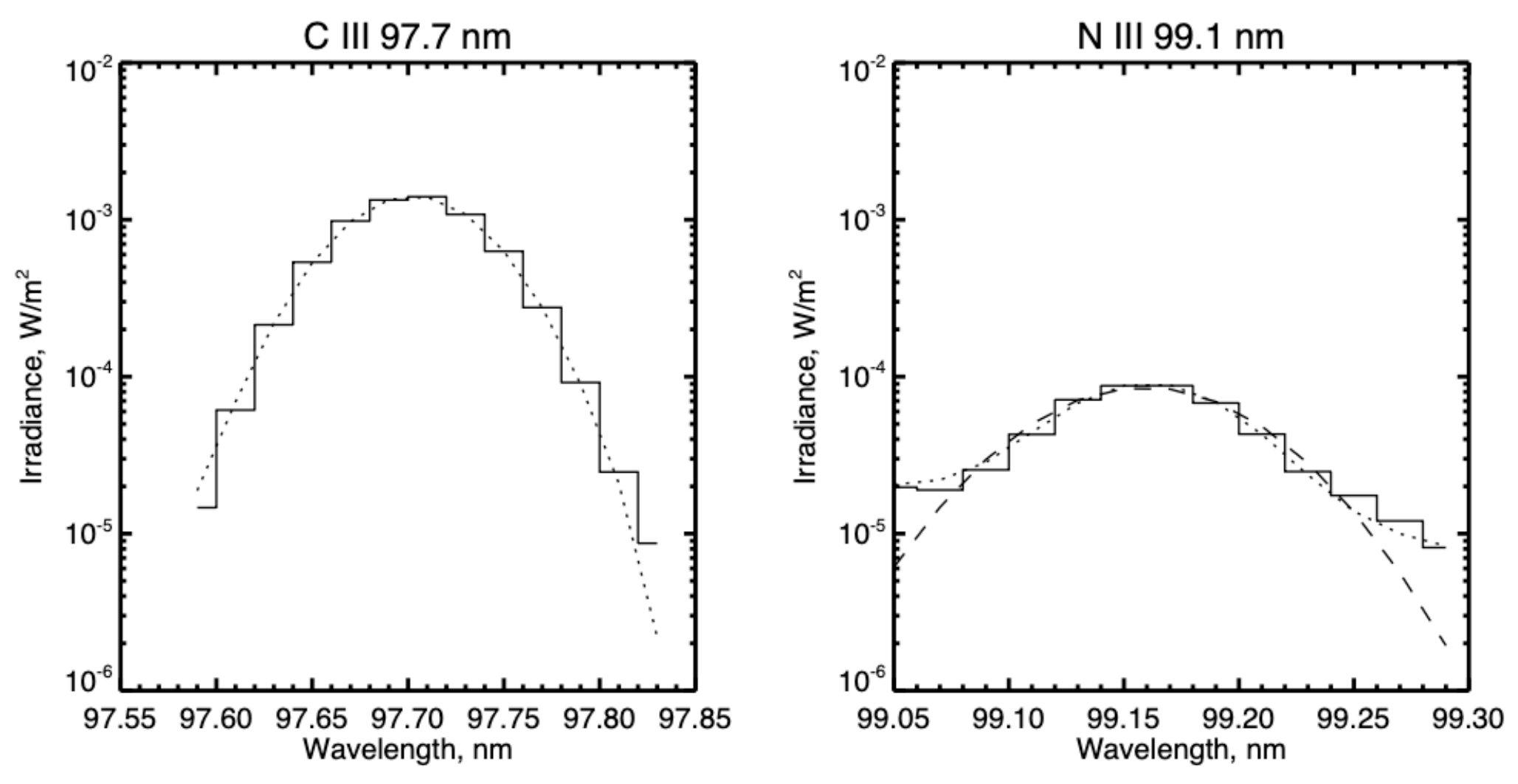}
\caption{\textit{Examples of individual EVE 10-s spectra, the C~{\sc iii} 97.7~nm and N~{\sc iii} 99.1~nm lines recorded 1-Jan-2011 at 00:50:10~UT (histograms).
Standard Gaussian fits (dotted line, 6-parameter; dashed line, 3-parameter) to these lines yielded the parameters (median wavelengths) listed in Table~\ref{tab:double_fit} (see text for discussion).
These are typical of the 360,000 fits used to characterize the time-series behavior of each line over the year 2011.}
\label{fig:double_fit}
}
\end{figure}

The basic data reduction proceeds from the EVE spectra; for each line and each 10-s sample we fit a single Gaussian plus a background quadratic term.
Figure~\ref{fig:double_fit} shows representive examples.
Note that all data used in this paper are the Level-2 data product\footnote{File of type \textit{EVS\_L2\_2011001\_00\_008\_01.fit}.}, and that the data reduction to this level results in a simplified Gaussian-like line profile with uniform properties.
This means that we cannot directly study line widths, but the line centroids do retain the excellent precision that the lower panel of Fig.~\ref{fig:feb_timeseries} illustrates.
We use IDL's {\sc fft\_powerspectrum.pro} to analyse fluxes and wavelengths obtained from the Gaussian fits.
Figure~\ref{fig:double_fit} shows 10-s spectra with the standard 6-parameter Gaussian fit (a Gaussian plus a quadratic ``background'' term). 
For the blended N~{\sc iii} 99.1~nm line we also show a simple 3-parameter Gaussian as an illustration.
The wavelength centroid of the fit can be determined much more precisely than the width of a single 0.1~nm EVE wavelength bin (see Table~\ref{tab:double_fit}).

The EVE data in 2011 contain daily three-hour contiguous intervals.
We compute the power spectrum for each of these data chunks and then average them to obtain a single spectrum with good noise properties, thus limited to a frequency resolution of about 0.1~mHz; at this resolution the p-mode peaks blur together, but we can integrate over the envelope to determine the power in the p-mode band, which we define as 1--5~mHz.
Similar average-spectrum analysis for HMI was shown in the example of Figure~\ref{fig:hmi_fft},
where a distinct p-mode response clearly appears; HMI observations were continuous, so we could also use 2-day chunks that begin to resolve the individual p-mode peaks.

\begin{figure}
\centering
    \includegraphics[width=\linewidth]{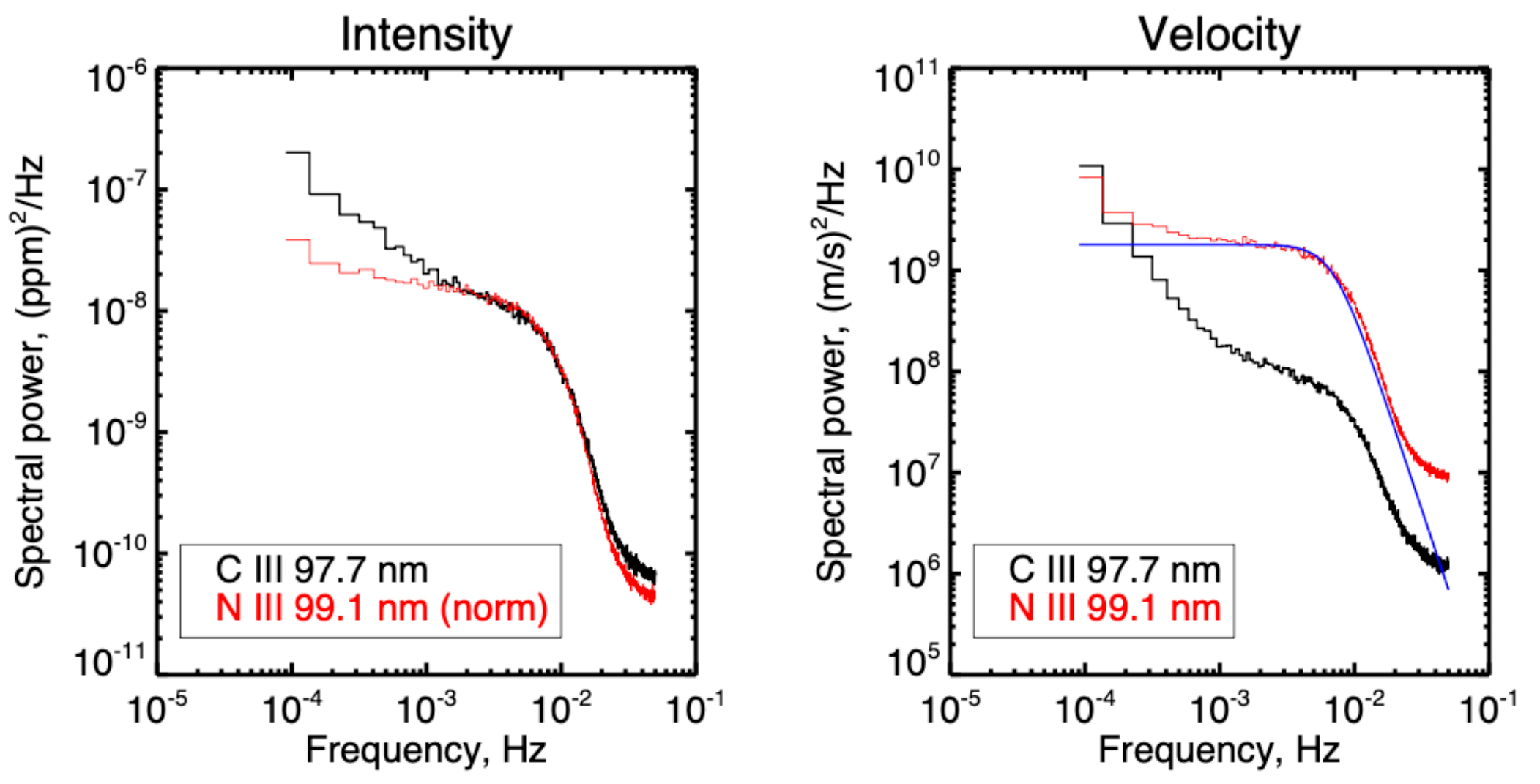}
\caption{\textit{In black, the average power spectra for a total of 231 daily 3-hr sequences of C~{\sc iii} 97.7~nm data in 2011, for intensity (left) and Doppler redshift (right). 
The red histograms show N~{\sc iii} 99.1~nm. 
The G(T) functions for these strong lines (see Figure~\ref{fig:master_goft}) coincide closely for peak G(T), but there is a high-temperature extension for the C~{\sc iii} and a greater OI value (Table~\ref{tab:linelist}). 
The blue line shows a Harvey function (Equation~\ref{eq:harvey}) fitted to the N~{\sc iii} Doppler spectrum, leading to the interpretation discussed in the Section~\ref{sec:turb}.}
}
\label{fig:ciiiniii}
\end{figure}

Figure~\ref{fig:ciiiniii} shows summed power spectra of both intensity and line centroid (shown as Doppler redshift), for the entire year 2011, for the strong transition-region lines C~{\sc iii} 97.7~nm and N~{\sc iii} 99.1~nm, whose G(T) functions are shown as heavy lines in Figure~\ref{fig:master_goft}.
We use the high-frequency intensity fluctuations (in the ``noise band'' $45-50$~mHz) in each chunk to identify problematic individual spectra; this reduced the number of usable chunks from 335 to 231 for both C~{\sc iii} and  N~{\sc iii}.
The power spectra using residuals against the median for each data chunk, without detrending the timeseries or applying a window function.
The FFT code requires uniformly-spaced data, but some 10-s data samples result in poor fits and thus gaps.
We have set those points to the median, likely resulting in a minor extra amount of high-frequency noise.
Finally, we have corrected the Doppler timeseries to eliminate the unwanted diurnal pattern due to SDO orbital motion (see Fig.~\ref{fig:feb_timeseries}) via ephemeris data. 
In the left panel of Fig.~\ref{fig:ciiiniii}, we have adjusted the intensity power spectrum for the N~{\sc iii} line by normalising it to match at 5~mHz.

\begin{table}
\caption{Representative line fits}\label{tab:double_fit}
\begin{tabular}{ l l l l}
\hline
Line & C~{\sc iii}  &  N~{\sc iii} & N~{\sc iii} \\
Parameters & 6 & 6 & 3 \\
Centroid (nm) &97.7025  &   99.1616   &  99.1592 \\
Error (nm) &0.00015  & 0.00101  & 0.00254 \\
\hline
\end{tabular}
\end{table}

\subsection{Line blends}\label{sec:blends}

The relatively low resolution of the EVE spectra means that there will be overlapping lines (blends).
We have selected a set of relatively unblended lines as listed in Table~\ref{tab:linelist}.\footnote{For a full discussion of EVE line spectroscopy, see \cite{2013A&A...555A..59D}.
}
The chosen set includes the higher Lyman series ($\beta, \gamma, \delta$), a series of oxygen ions (ionisation states {\sc ii-vi}), Mg ions {\sc vii-x}, Ne lines to provide first ionization potential (FIP) sensitivity, and the strong transition-region lines of C~{\sc iii} (97.7~nm) and N~{\sc iii} (99.151~nm).
The table provides metadata for the selected lines, using atomic data from CHIANTI \citep{1997A&AS..125..149D}, and Figure~\ref{fig:master_goft}.
Note that all lines in the Table generally yielded good Gaussian fits, even where blends were noticeably present.
The OI (``opacity index'') in the table is log$_{10}(AG-27)$, where $A$ is the photospheric 
abundance and $G$ is the maximum value of the $G(T)$ contribution function.

\begin{table}
\caption{Spectral line list}\label{tab:linelist}
\begin{tabular}{ l r c c l l l l}
&Ion& $\lambda$ & log~\textit{T} & $\rho^a$ & OI$^d$ & Note & L4L$^e$ \\
\hline
 H  &    I  &   1025.72  &  4.50 & 1.13 & 3 & Blend? & Clean  \\
 H  &    I  &    972.54  &  3.84 & 1.12 & 3 & Clean  & Clean      \\
 H  &    I  &    949.70  &  3.84 & 1.12 & 2 & Blend? & Clean \\
 He &    I  &    537.03  &  4.50 & 1.03 & 2 & Blend!! & N/A     \\
 He &    I  &    584.34  &  4.50 & 1.04 & 3 & Clean   & Clean     \\
 O  &   II  &    718.50  &  4.75 & 1.01 & 3 & Blend?  & N/A      \\
 O  &  III  &    835.50  &  4.95 & 1.38 & 2 & Clean$^b$ & Blend!!    \\
 O  &  III  &    525.79  &  5.02 & 1.06 & 3 & Blend   & Clean     \\
 O  &  III  &    599.59  &  5.02 & 1.05 & 4 & Blend?  & Clean      \\
 O  &   IV  &    554.51  &  5.22 & 1.20 & 4 & Blend   & Clean     \\
 O  &   IV  &    790.20  &  5.20 & 1.40 & 4 & Blend!  & Clean     \\
 O  &    V  &    629.73  &  5.39 & 1.08 & 5 & Clean   & Clean     \\
 O  &   VI  &   1031.91  &  5.50 & 1.00 & 4 & Clean   & Clean      \\
Mg  &  VII  &    434.92  &  5.80 & 1.21 & 2 & Blend!! & N/A     \\
Mg  & VIII  &    436.73  &  5.99 & 1.14 & 3 & Blend?  & N/A     \\
Mg  &   IX  &    368.07  &  6.00 & 1.01 & 3 & Clean   & Clean      \\
Mg  &   IX  &    439.18  &  6.00 & 1.33 & 1 & Blend!! & N/A    \\
Mg  &    X  &    609.80  &  6.07 & 1.00 & 3 & Clean   & Clean     \\
Mg  &    X  &    624.94  &  6.07 & 1.00 & 2 & Clean   & Clean      \\
Ne  &  VII  &    465.22  &  5.72 & 1.02 & 4 & Clean   & Clean     \\
Ne  & VIII  &    770.41  &  5.81 & 1.00 & 3 & Clean   & Clean     \\
Si  &  XII  &    499.41  &  6.29 & 1.00 & 3 & Blend?  & Clean     \\
Si  &  XII  &    521.00  &  6.29 & 1.00 & 2 & Blend   & Clean     \\
C   &  III  &    977.16  &  4.94 & 3.14 & 4 & Clean   & Clean     \\
N   &  III  &    991.51  &  4.94 & 1.83 & 2 & Blend!!$^c$   & N/A    \\
Fe  &  XVI  &    335.41  &  6.50 & 1.00 & 3 & Blend?  & Clean     \\
\hline\noalign{\smallskip}
\end{tabular}

$^a$ G(T) peak ratio for assumed $n_e$ values (cm$^{-3}$) 10$^{11}/10^{9}$.\\
$^b$ Heavily blended in EVE by other O~{\sc iii} lines.\\
$^c$ Heavily blended in EVE by other N~{\sc iii} lines.\\
$^d$ An ``opacity index'' combining G(T) and abundance logarithmically (see text).\\
$^e$ \cite{2026SoPh..301...63W}.
\end{table}

Table~\ref{tab:linelist} describes the appearance of each of the 25 lines studied.
Of these 10 appear subjectively to be clean, with the others potentially blended.
In the Notes column, ``Clean'' means no significant quadratic term in the standard 6-parameter Gaussian fit and less likelihood of a blend.
More obvious blends are listed as ``Blend!'' or ``Blend!!'' if very distinctive; this grading is semi-quantitative based on the prominence of the fitted line relative to the quadratic background term.
\cite{2026SoPh..301...63W} provide a more authoritative assessment of each of these lines with the EVE level~4 data product (``L4L''), which includes direct fitting of multiple lines as needed.
In our parallel assessment, we take lines with enhanced quadratic ``background'' levels, as judged across our 13-point (2.4~\AA) analysis window, to reflect the presence of weak lines.
We have basically ignored these qualifications, relying on the single-Gaussian profile fitting, and have ultimately found no obvious inconsistencies with the results described below.
The two blend assessments usually agree, but for 6/25 lines the L4L judgement is ``N/A'', meaning very blended.

\begin{figure*}
\centering
    \includegraphics[width=0.24\textwidth]{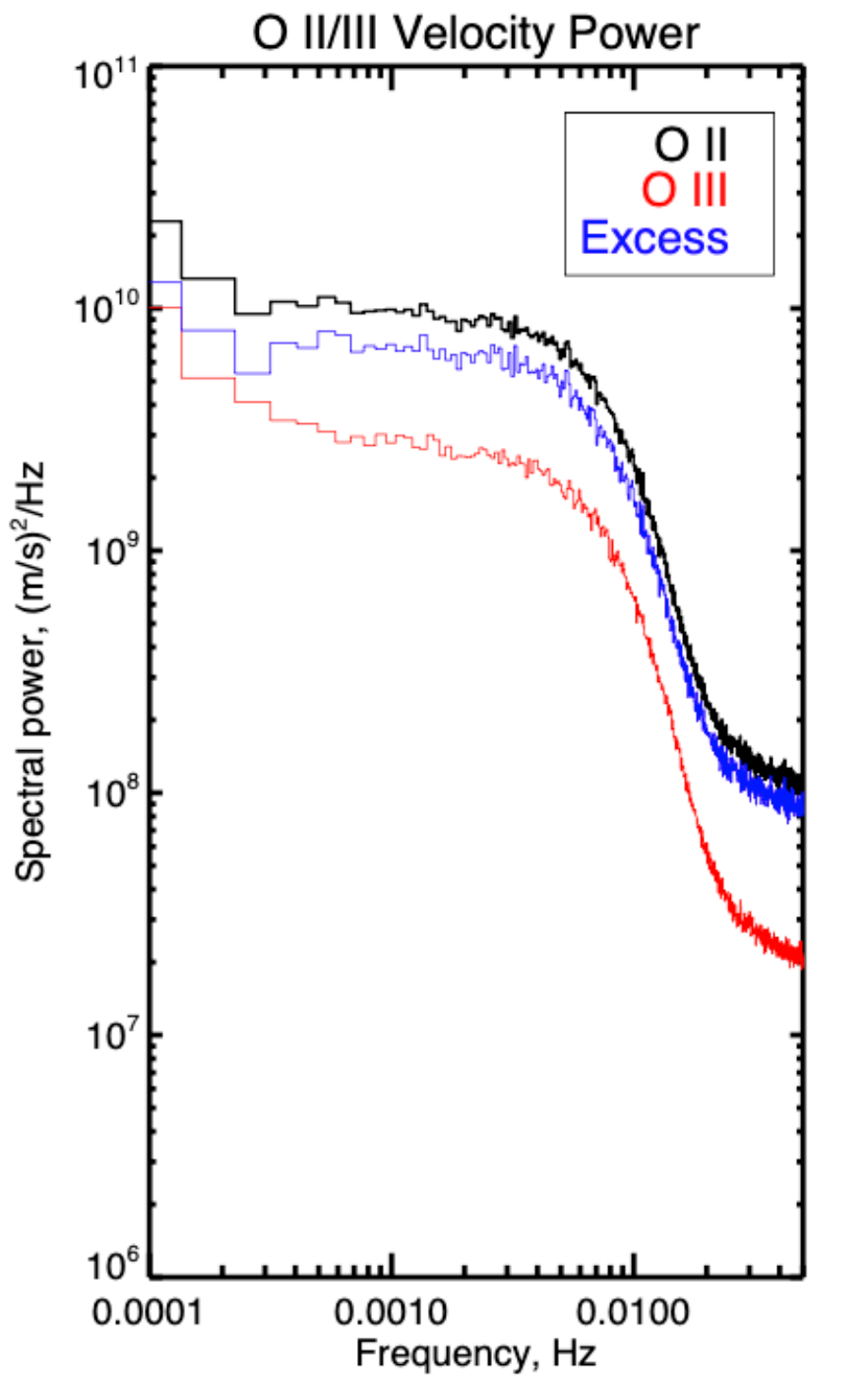} 
    \includegraphics[width=0.24\textwidth]{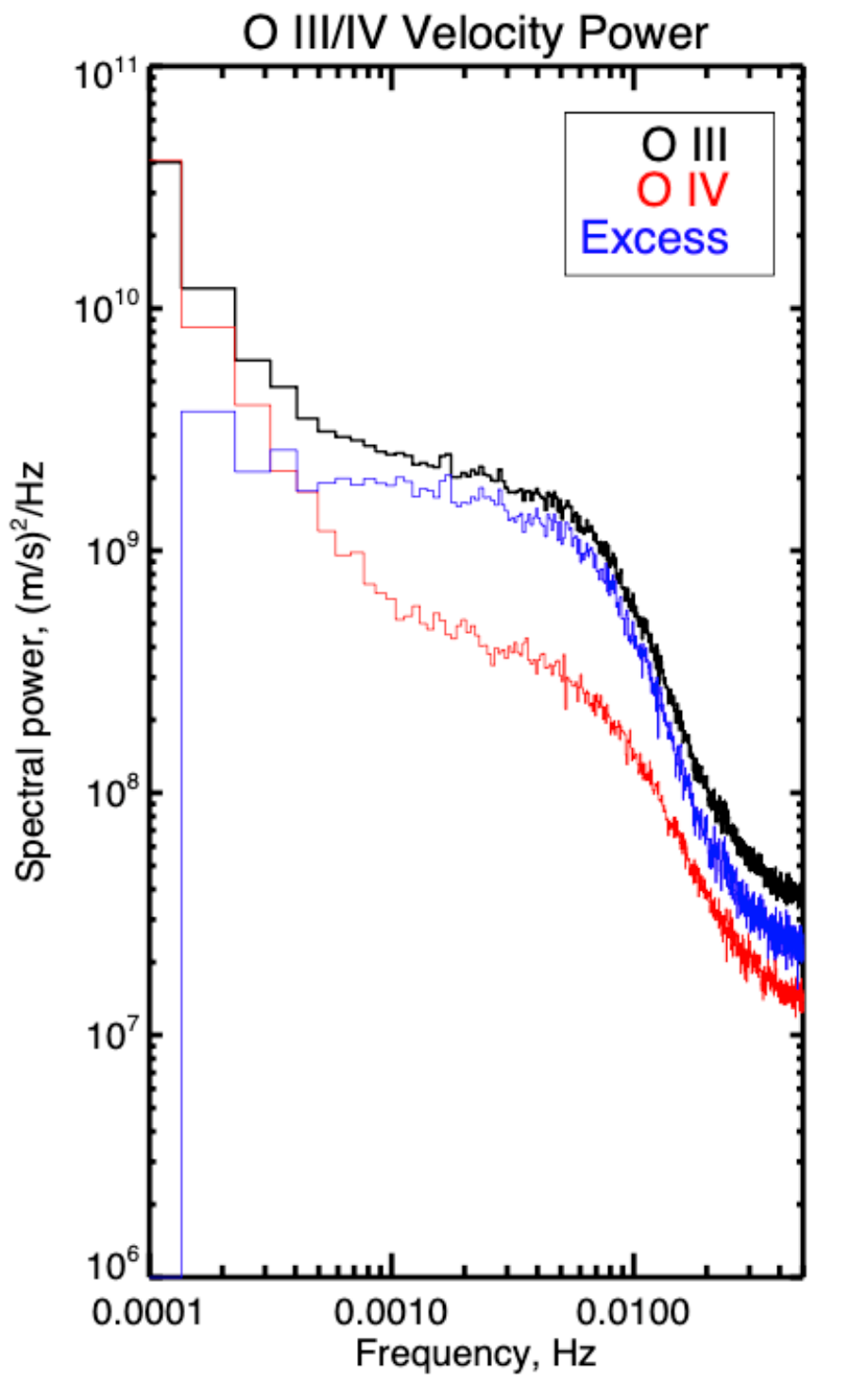} 
    \includegraphics[width=0.24\textwidth]{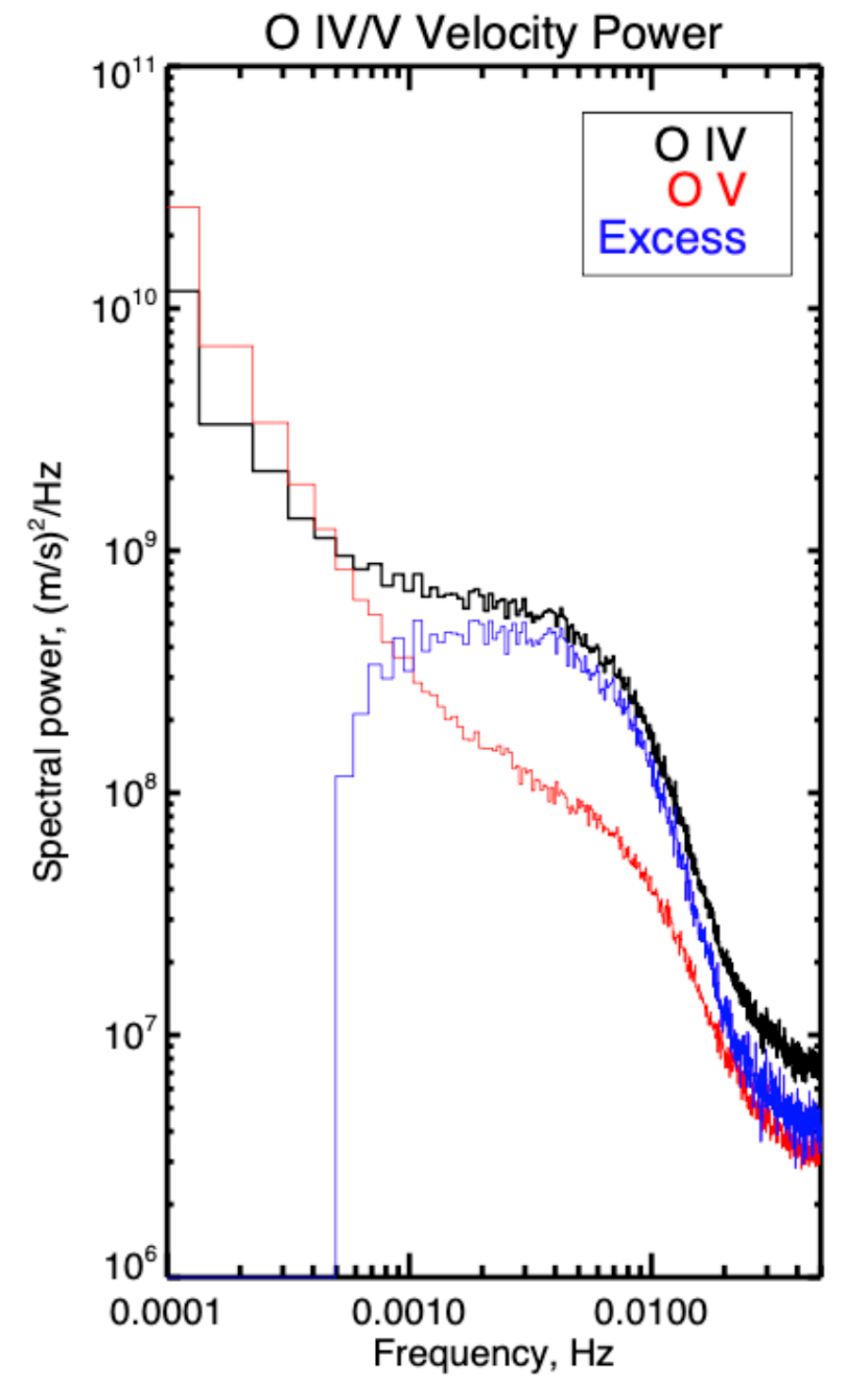} 
    \includegraphics[width=0.24\textwidth]{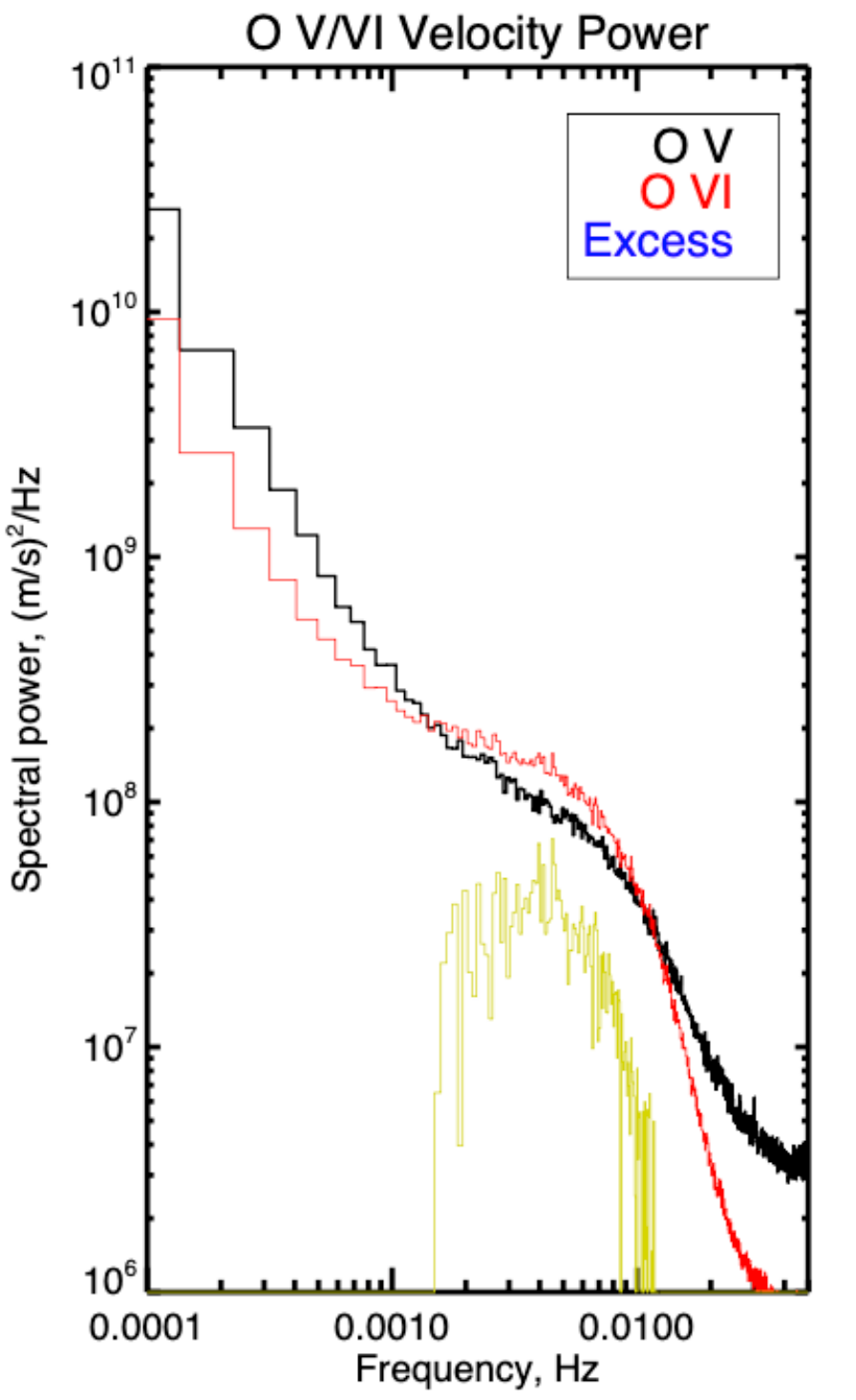} 
     \caption{\textit{Power spectra from adjacent pairs of O ions, ranging from O~{\sc ii} to  O~{\sc vi} and ignoring blend warnings.
     The difference spectrum (gold color) in the right panel is opposite from that of the first three pairs (blue), which clearly establish an increasing suppression of the few-mHz power relative to the lower frequencies.}
        }  
\label{fig:o_lines} 
\end{figure*}

We believe that we can generally ignore line blends, as discussed above. 
Figure~\ref{fig:double_fit} and Table~\ref{tab:double_fit} further illustrate this point; the comparison of 6- and 3-parameter fits for  the heavily blended N~{\sc iii} line shows tolerable absolute centroid displacements, and suggest negligible effects on the variances shown in our power-spectrum analysis.
Figure~\ref{fig:o_lines} confirms this empirically by showing an orderly progression through the oxygen ionization states, {\sc ii-vi}, despite using lines with the various blend assessments listed in Table~\ref{tab:linelist}.

The Table also checks for density sensitivity, comparing CHIANTI G(T) maxima for test densities of 10$^9$ and 10$^{11}$~cm$^{-3}$ (column 5).
Column~6 is the ``opacity index'' defined above, a logarithmic product of peak G(T) and (photospheric) abundance.

\subsection{Morphology}

The data give us very clean views of mean solar variability up to 50 mHz that cannot be well modeled at present,  owing to EVE's Sun-as-a-star viewpoint and the need for long integration time.
Figure~\ref{fig:master_goft} shows that EVE has broad temperature coverage, but we cannot translate that into geometrical formation heights in any quantitative way.
Our Sun-as-a-star view combines the whole range of atmospheric heights, averaged over an entire year, with no clear guidance to the effects of optical depth (limb darkening) nor image structure.

We nevertheless can draw some conclusions by inspection of these spectra.
The higher frequencies available from the EVE 10-s sampling do not show a white-noise floor in the power spectra of either intensity or radial velocity just below the Nyquist frequency of the 10-s sampling, 50~mHz.
This implies high signal-to-noise ratio in the individual 10-s integrations.
The standard approach for describing Sun-as-a-star power spectra is that proposed by \cite{1985ESASP.235..199H,1993ASPC...42..111H}, which uses one or more Lorentzian-like components to represent different scales of convective motions, as in Equation~\ref{eq:harvey}.

\begin{equation}
\label{eq:harvey}
P(f) = A f^{-n} + \frac{B}{1+(f/f_0)^\alpha} + C \ \ .
\end{equation}

The power spectra (in units of $(m/s)^2/Hz$ as a function of frequency $f$) resemble those expected from shot noise: flat at low frequencies, rolling over to a 1/$f^2$ form above a break frequency of $1/\tau$ for a characteristic time scale $\tau$.
The solar driver does not consist of discrete shots, but rather of smoother and more
continuous variations.
The formula describes this with a 1/$f^\alpha$ trend at high frequencies, and we discuss the interpretation of this form as turbulence in Section~\ref{sec:turb}.
A generalized Harvey-type spectrum can describe visible/TSI spectra well \citep{2021ApJ...916...66B}, with  A, B, C, $\alpha$ and $f_0$ the fitted parameters for a single scale of convection.
Multiple scales of convection, for instance, the supergranulation at larger spatial scales and lower frequencies, can have the same distribution and add to the total spectral power.
\cite{2012MNRAS.421.3170K} also detected a facular scale in this manner.
\cite{2021ApJ...916...66B} interestingly suggest that the supergranulation term in the VIRGO photometric data may result from bright points in the network, rather than large-scale flows.
At still lower frequencies, on solar rotational time scales, an activity-dependent term appears in the TSI power spectra \citep{2004A&A...414.1139A}.
Frequencies below about 10~$\mu$~Hz (a day) contain signatures of magnetic activity and stellar rotation, which is beyond the scope of this paper.

The Harvey function clearly captures the nature of the spectra: a roughly power-law continuum with an excess coinciding with typical time scales for granulation.
We have instead simply characterized the spectra into five regions: $0.2-1$~mHz, 1-5~mHz, 5-20~mHz, 20-45~mHz, and 45-50~mHz (Table~\ref{tab:bands}).
These would roughly correspond to supergranulation, the p-mode range, a high-frequency range, a ``tail range', and a noise-level band respectively.
In practice, we do not detect the p-modes individually (see below);  they should be only weakly detectable in the EUV if at all, so one should consider the entire range 1-20~mHz as due to convectively-driven motions in the scale range of the granulation.

\begin{table}
\caption{Frequency bands}\label{tab:bands}
\begin{tabular}{ l l}
\hline
Name & Frequencies \\
     & mHz \\
\hline
all & 0.1-50 \\
super & 0.2-1 \\
p-mode & 1-5 \\
high-f & 5-20 \\
tail & 20-45 \\
noise & 45-50 \\
\hline
\end{tabular}
\end{table}

\begin{figure}
\centering
    \includegraphics[width=\linewidth]{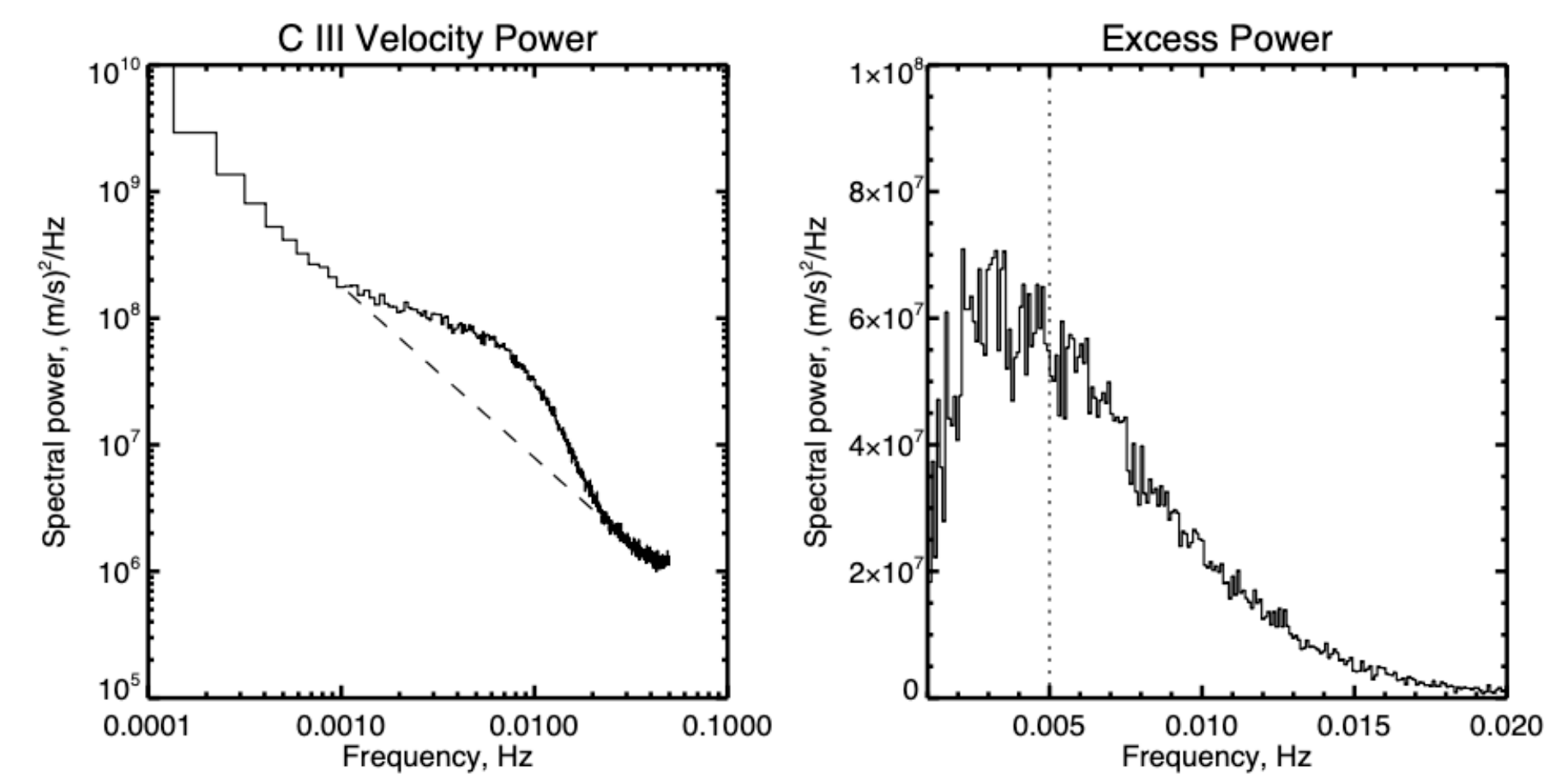}    
    \includegraphics[width=\linewidth]{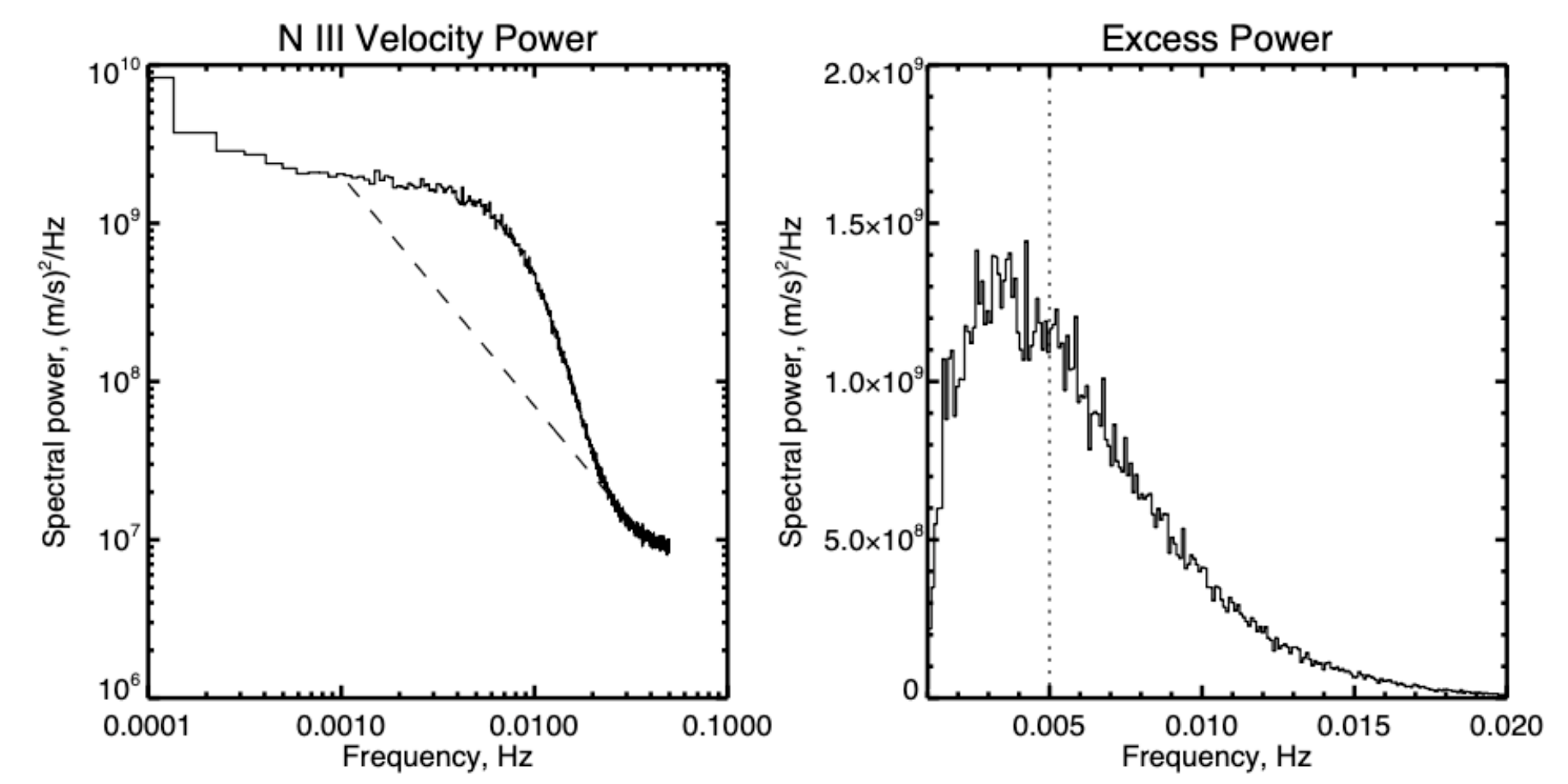}
\caption{\textit{Incoherent-sum power spectra for C~{\sc iii} 97.7~nm (upper) and N~{\sc iii} 91.1~nm (lower).
Left panels, the total power spectra; right panels, the excess power levels above a power-law interpolation between 1-30~mHz, on a linear scale.
The vertical dotted line at 5~mHz separates the p-mode band from the higher-frequency continuum, due mainly to the Harvey spectrum of convection on granulation scales.
}
}
\label{fig:ciii_2up}
\end{figure}

We can roughly isolate the spectral Doppler power at granulation scales by subtracting a power-law continuum, interpretable as the high-frequency tail of the spectral power due to supergranulation.
Figure~\ref{fig:ciii_2up} shows the excess Doppler power using an \textit{ad hoc} power-law fit over 1--30~mHz for the background term.
The vertical dotted line separates the p-mode region from the higher frequencies of the Harvey function for the granulation scales.
We do not expect to detect p-modes directly, a topic discussed further below, but note here that the continuum below 5~mHz does not resemble the HMI p-mode envelope seen in the photosphere (Figure~\ref{fig:hmi_fft}).
The background-subtracted power levels in the broad bands do not differ greatly from the direct integrations, and so we do not make further use of them.

This article mainly describes the Doppler power, but for completeness we also show power spectra of representative line fluxes in Section~\ref{sec:flux}.
Flux and Doppler spectra basically have the same morphology.

\subsection{The p-mode band}

\begin{figure}
\centering
     \includegraphics[width=0.8\linewidth]{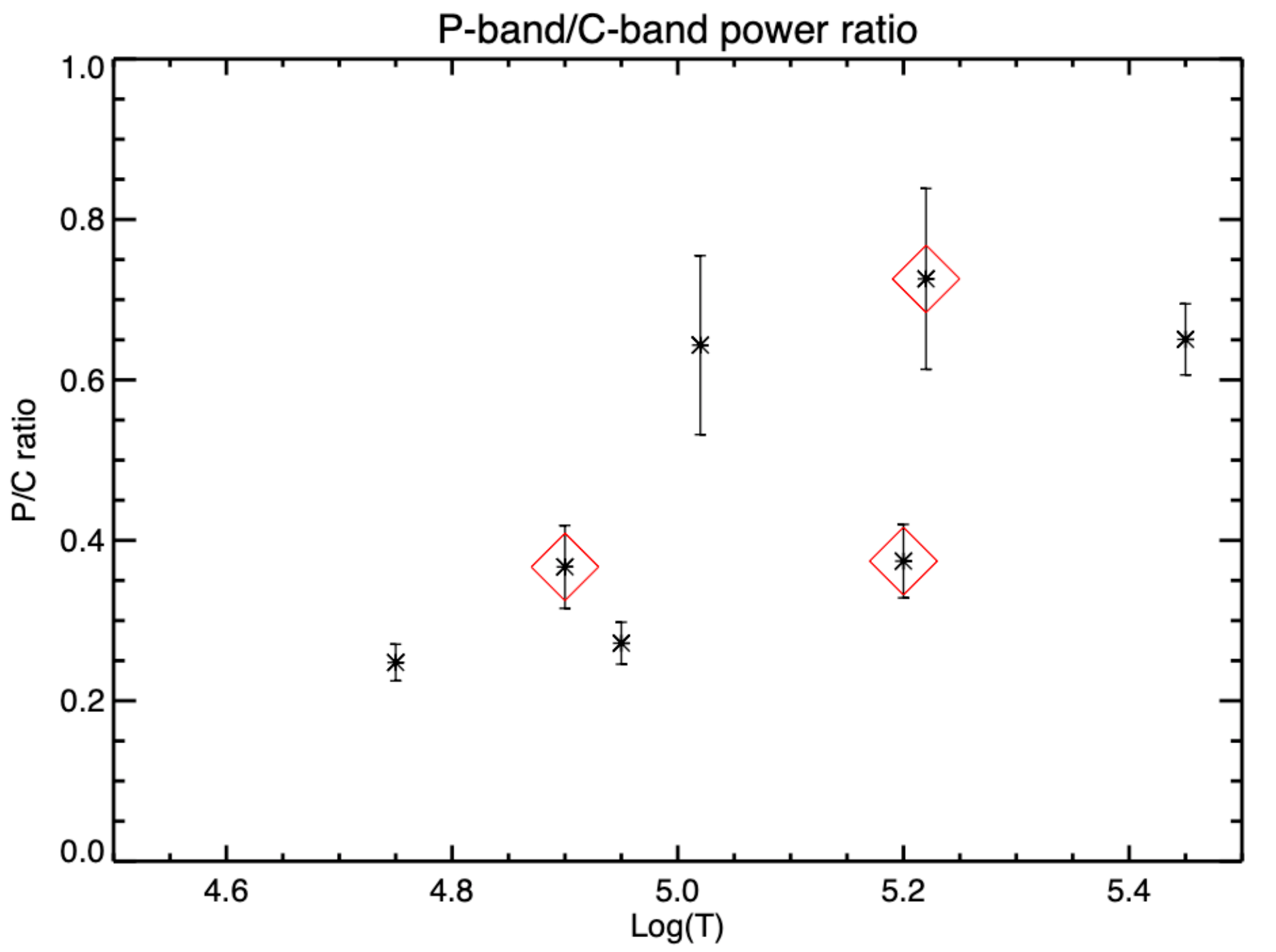}
\caption{\textit{Ratio of p-band Doppler power (1-5~mHz) to c-band power (convective continuum, 5-20~mHz) for various ionisation states of O, as sorted by the CHIANTI log(\textit{T}) [K] values.\
The points highlighted in red represent lines with some blending.}
}
\label{fig:pbandcband}
\end{figure}

We have crudely searched for the possible p-mode power contributions within the oxygen line series by comparing the Doppler power levels in our p-band (1-5~mHz) and c-band (5-20~mHz).
We could expect these contributions to vary with height in the atmosphere, and therefore be reflected in the temperature of line formation, as approximated by the ionization state.
Figure~\ref{fig:pbandcband} shows the ratio of powers for our sequence of oxygen ions (Table~\ref{tab:linelist}, showing some dependence on the ionization state but also revealing some unexplained variance).
This crude measure is not satisfactory, given the excess scatter of the data points, but the blending does not seem to play a role.
At this frequency resolution we could not resolve individual p-mode peaks in any case and believe that a higher-resolution search would be feasible and interesting, noting that the Sun-as-a-star view will include local regions where special conditions for wave propagation may arise \citep[e.g.,][]{2026ApJ...999L...4H}; see also \cite{2008JGRA..113.8110B}.
The positive correlation seen in Figure~\ref{fig:pbandcband} probably does not imply anything about p-mode amplitudes, but it does hint at unexplained systematic errors.

\subsection{Spectral flux}\label{sec:flux}
\begin{figure}
\centering
    \includegraphics[width=0.49\linewidth]{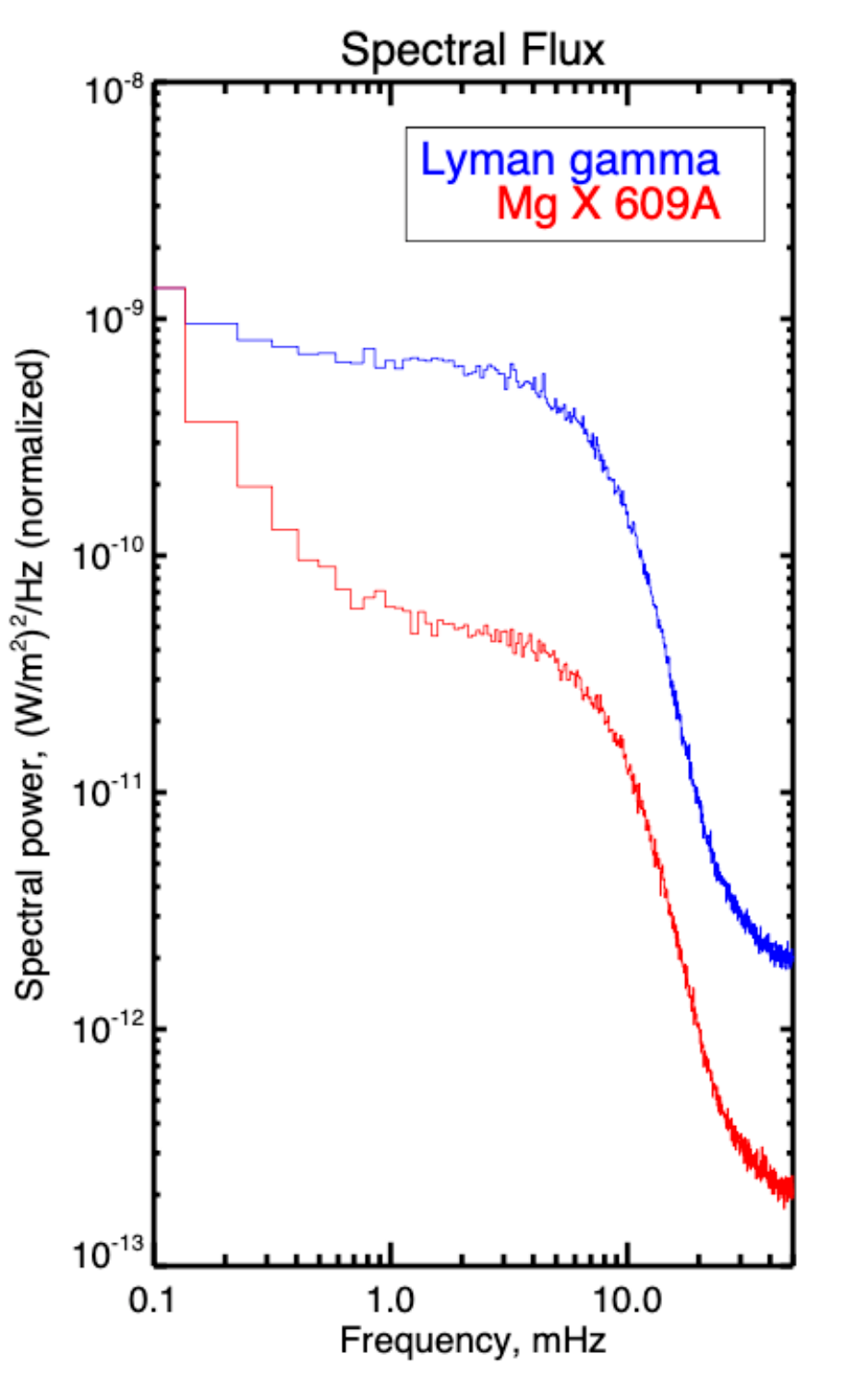}    
    \includegraphics[width=0.49\linewidth]{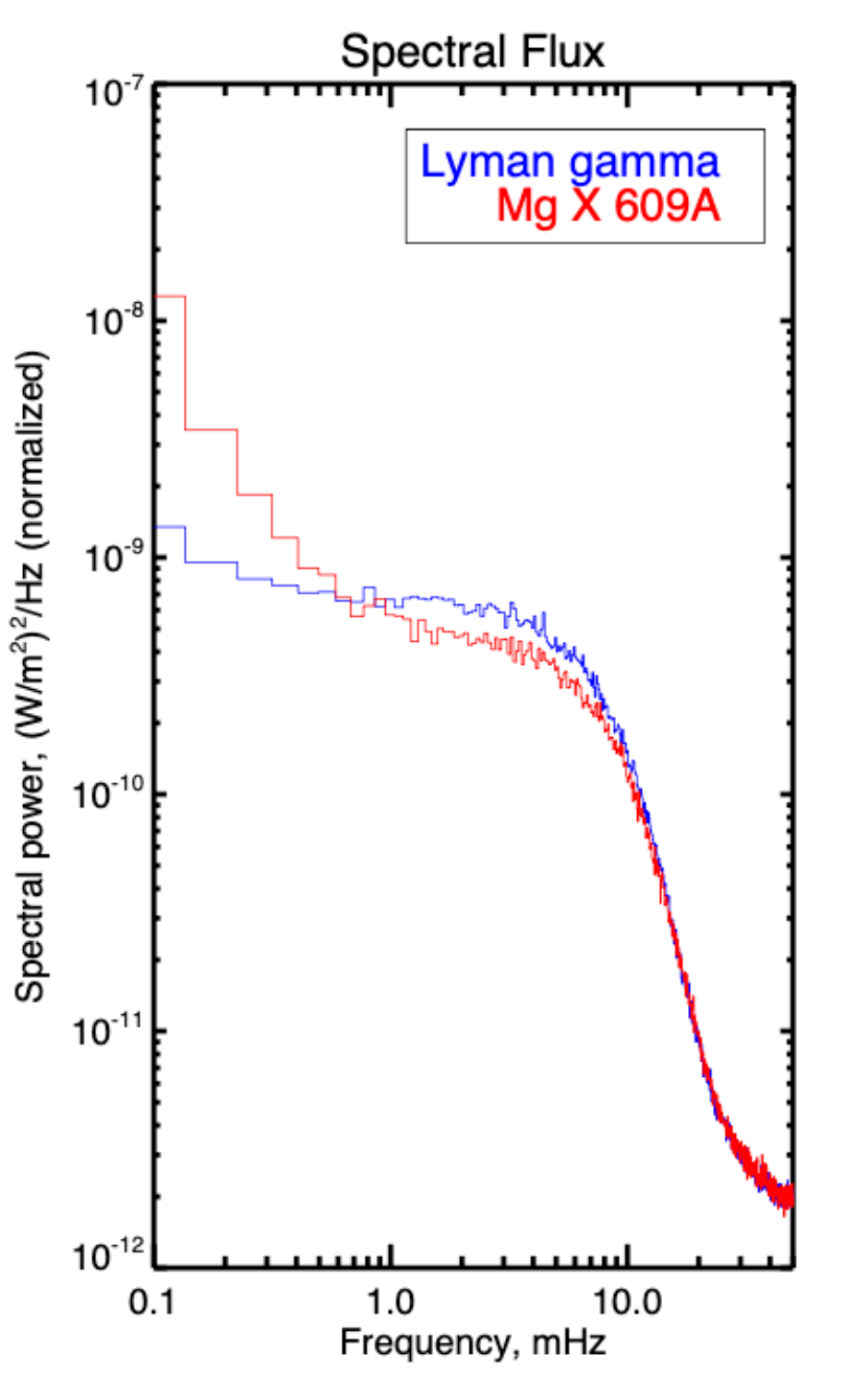}
\caption{\textit{Comparison of spectral power (flux, not Doppler) for representative hot and cold lines, with the left panel showing normalization to the Lyman-line spectrum at 1~mHz, and the right panel to the noise band at 45-50~mHz.
These are both arbitrary in the sense that the spectra are given in dimensional units, and just serve to illustrate the 
relative excess of supergranulation-scale power in the hot line.}
}
\label{fig:hotcold}
\end{figure}

We focus here on Doppler spectra, but for comparison with spectral power distributions in other wavelength ranges (such as total solar irradiance) we also compute spectral fluxes.
Figure~\ref{fig:hotcold} compares Mg~{\sc x}~609~\AA\ (CHIANTI logT = 6.07) with Lyman gamma 972~\AA\ (3.84).
Both of these lines are unblended and have opacity index 3 (Table~\ref{tab:linelist}.
These distributions look much like the Doppler spectra and display the same dichotomy, with the hot line having
greater low-frequency power in the supergranulation time range.
We thus see the same morphology in both Doppler and flux spectra.

\subsection{The continuum Doppler spectrum and turbulence}\label{sec:turb}

The bulge at about 5~mHz clearly reflects a Harvey spectrum of motions on granulation scales.
We have chosen a set of values for parameters from Equation~\ref{eq:harvey} and show these as a blue line to the Doppler power spectrum for the C~{\sc iii} line in Figure~\ref{fig:ciiiniii}. 
This fit has a power-law index $\alpha \approx 4$ (Equation~\ref{eq:harvey}).
As discussed by \cite{1993ASPC...42..111H} and by \cite{2012MNRAS.421.3170K}, for smoothly varying convective motions at a specific time scale, the high-frequency tail of the bump power spectrum will decay exponentially with frequency. 
Our spectra do show a steep fall-off in the 10-20~mHz range, but one closer to a power law than to an exponential decay.
The relatively high Nyquist frequency of the EVE data makes it possible to obtain good parameter fits for many lines, both in flux and Doppler spectra, and we can follow this power law for more than two decades.

The extension to 50~mHz in these power spectra in a steep power-law form does not resemble that expected from Kolmogorov turbulence, which would have an $f^{-5/3}$ dependence \citep{1994ApJ...432..612S}.
We can thus set upper limits on the amplitude of a Kolmogorov component by fitting a 5/3 law to the n-band (45-50~mHz) levels (``n'' for ``noise'').
The spectra at the Nyquist frequency flatten out, as is typical in such spectra limited by white noise.
In the case of the EVE spectra, this could readily be explained by spacecraft pointing jitter, for example, or photon statistics.
An underlying Kolmogorov spectrum of turbulence would be another
possibility, as would solar high-frequency power from network-scale magnetic activity, possibly contributing another Harvey function.
We do not think that N-band power contains much aliased power from frequencies above the Nyquist frequency.
The different levels of Doppler power in lines with different contribution functions indicates that the high-frequency power generally has a solar origin.
The shape of the Harvey function naturally might seem consistent with an outer scale of motions driving higher-frequency fluctuations in a Kolmogorov cascade, but this can be ruled out by the steepness of the high-frequency spectra.

\begin{figure}
\centering
    \includegraphics[width = 0.6\linewidth]{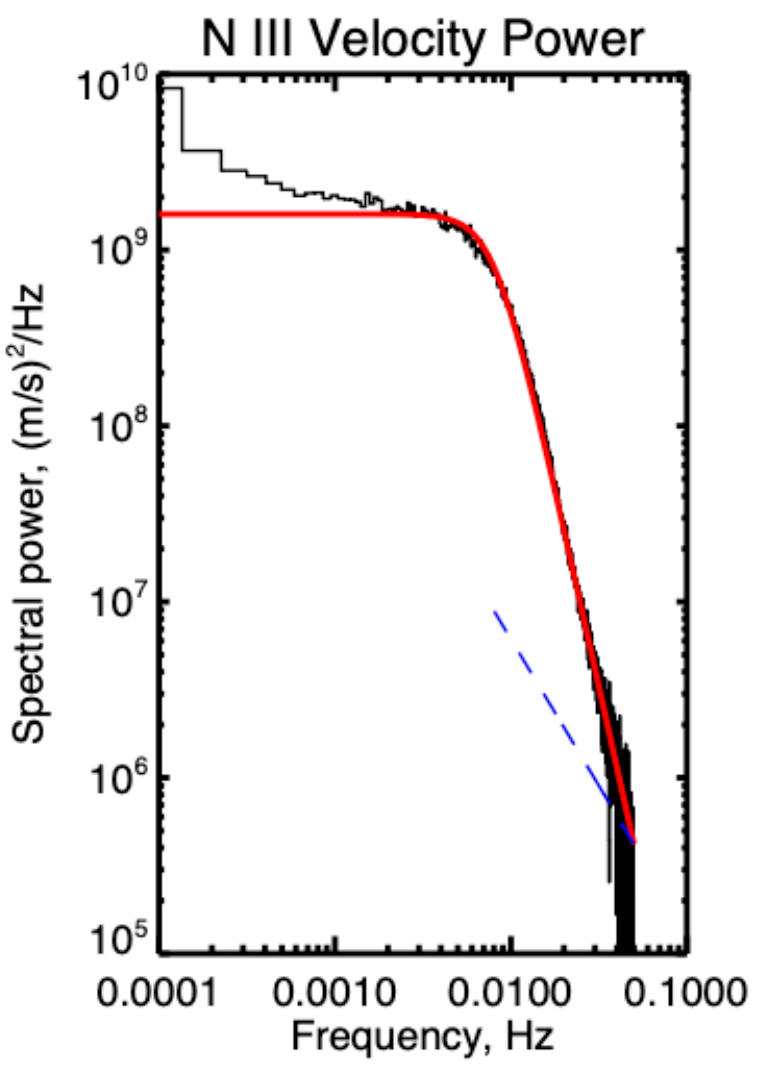}   
 \caption{\textit{Doppler spectrum for N~{\sc iii} 991.51~{\AA}, with a Kolmogorov spectrum normalized to the N-band power at 45-50~mHz shown as a blue dashed line.}
}
\label{fig:turb_limit}
\end{figure}

As a representative limit on the turbulent component, we show its upper-limit normalization in Figure~\ref{fig:turb_limit}.
The N~{\sc iii} line at 991.51~\AA\ forms at chromospheric temperatures (Table~\ref{tab:linelist}) and we assume that its Harvey spectrum defines the outer scale for a Kolmogorov component, shown as a blue dashed line. 
Here we have subtracted the n-band (45-50~mHz) power level, assuming it to represent white noise, and an \textit{ad hoc} fit to the Harvey function has a rollover frequency of 8~mHz, taken here to represent the outer scale.
The high-frequency power law in this case has $\alpha \approx 4.5$ (Equation~\ref{eq:harvey}).
We obtain an rough estimate for the ratio of turbulent variance to the total, $W_{turb}/W_{tot} \leq 10^{-5}$.
The spectrum represents an integration covering the whole Sun over an entire year.
No direct conclusion regarding specific solar features is possible, except to note that this limit applies basically to closed-field coronal regions.

From the point of view of spectroscopy, the levels that we detect contribute a fraction of the emission line widths, and likewise a  fraction of the frequently discussed ``non-thermal velocity'' $v_{nt}$ or ``microturbulence'' taken as an estimate of excess width in comparison with some assumed thermal width. 
The EVE Doppler results do not conflict with classical observations of emission line widths.
The well-studied visible red, green, and yellow lines show excess widths of 10-20~km/s \citep[e.g.,][]{1966gtsc.book.....B}, and Skylab results for several coronal UV lines \citep{1979ApJ...227.1037C} show similar magnitudes.
For comparison, our estimate including the low frequencies is $\sim$15~km/s rms, assuming isotropy.
For a full discussion of line widths in EUV spectroscopy, see \cite{2018LRSP...15....5D}.
The CoMP observations suggest that most of the excess line widths in quiescent coronal emission lines result from macroscopic flows 
\citep[e.g.][]{2007Sci...317.1192T}, not from fully-developed turbulence.
For comparison, our estimate including the low frequencies is $\sim$15~km/s rms, again assuming isotropy, and this would be a strong upper limit for the amplitude of coronal turbulence over a broad frequency band.

\subsection{Parameter dependence}\label{sec:parms}

\begin{figure}
\centering
    \includegraphics[width = 0.49\linewidth]{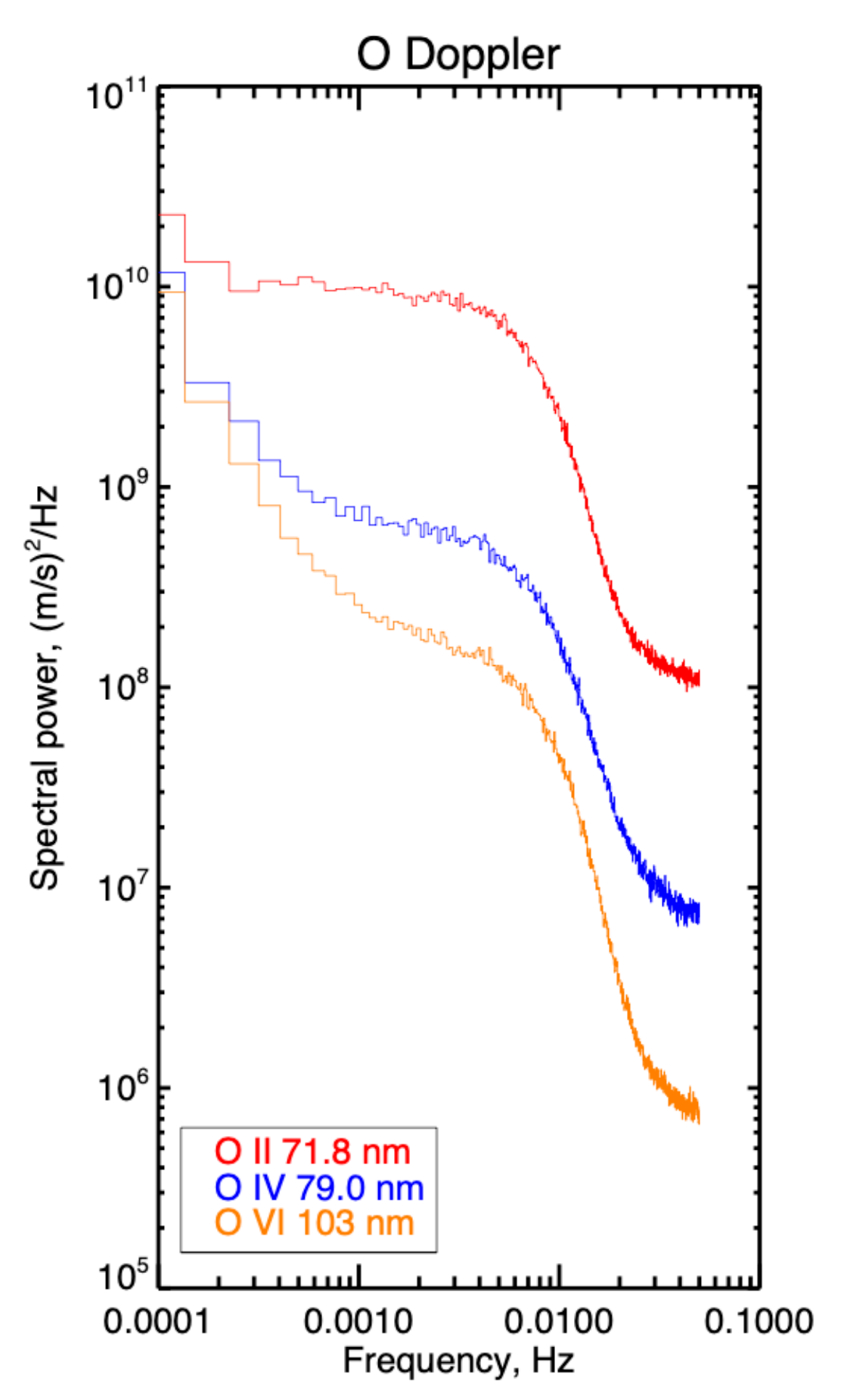}   
    \includegraphics[width = 0.49\linewidth]{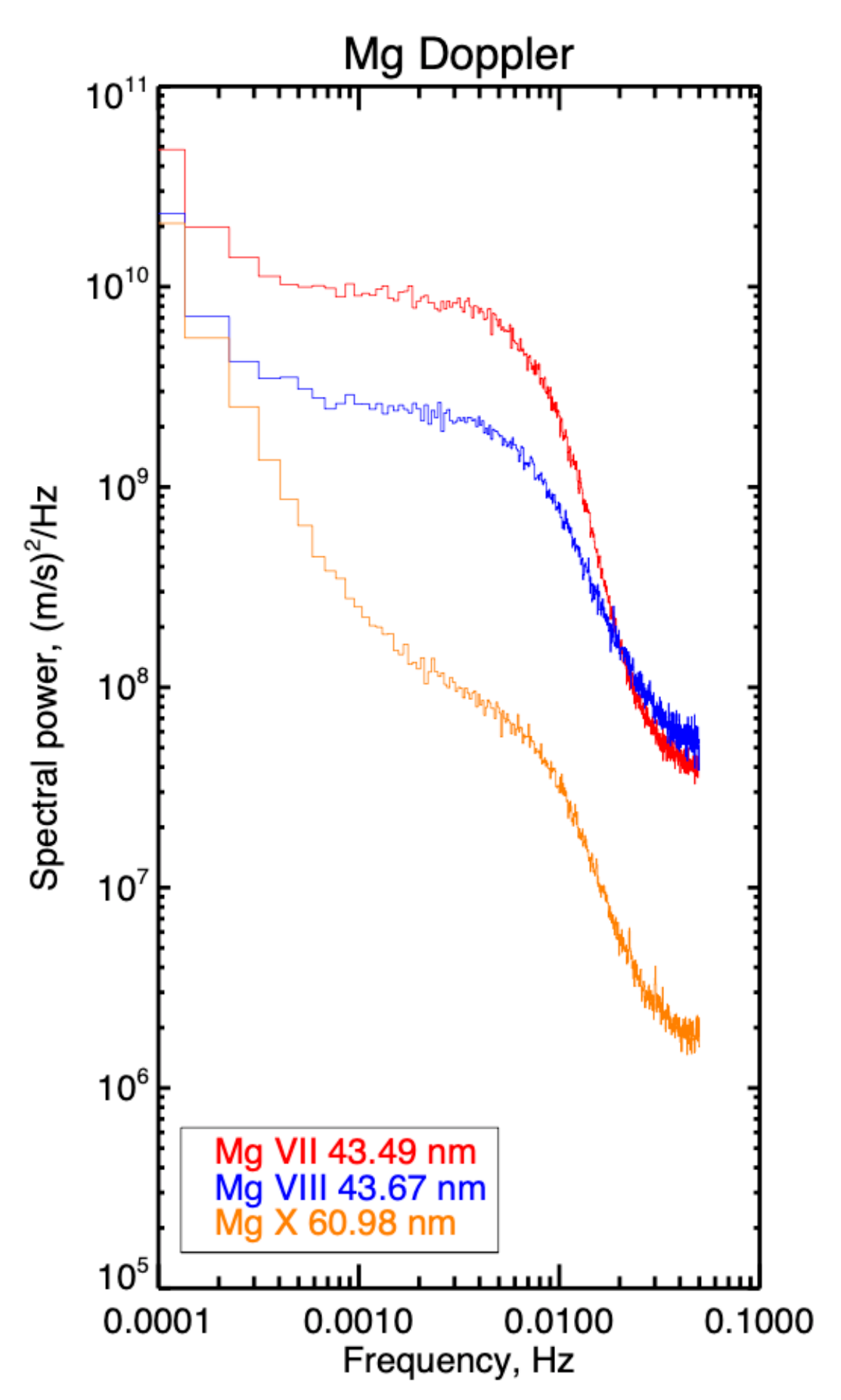}   
\caption{\textit{The Doppler spectra for three of the EVE ionization states each of oxygen and magnesium, with both sequences showing decreasing Doppler amplitude with temperature of line formation; note also the convergence to similar values a the lowest frequencies.}
}
\label{fig:OMg}
\end{figure}

The line list includes a set of five ionization states of oxygen ({\sc ii-vi}) and four ionization states (of magnesium ({\sc vii-x}), for which we show Doppler spectra in Figure~\ref{fig:OMg}.
The power levels differ substantially and systematically, revealing a relative decrease at higher frequencies as the line-formation temperature increases.
We can speculate that the granulation-scale power (the peak at 5~mHz) has greater concentration in the presumably smaller scales of the transition region and chromosphere.

\begin{figure}
\centering
    \includegraphics[width=\linewidth]{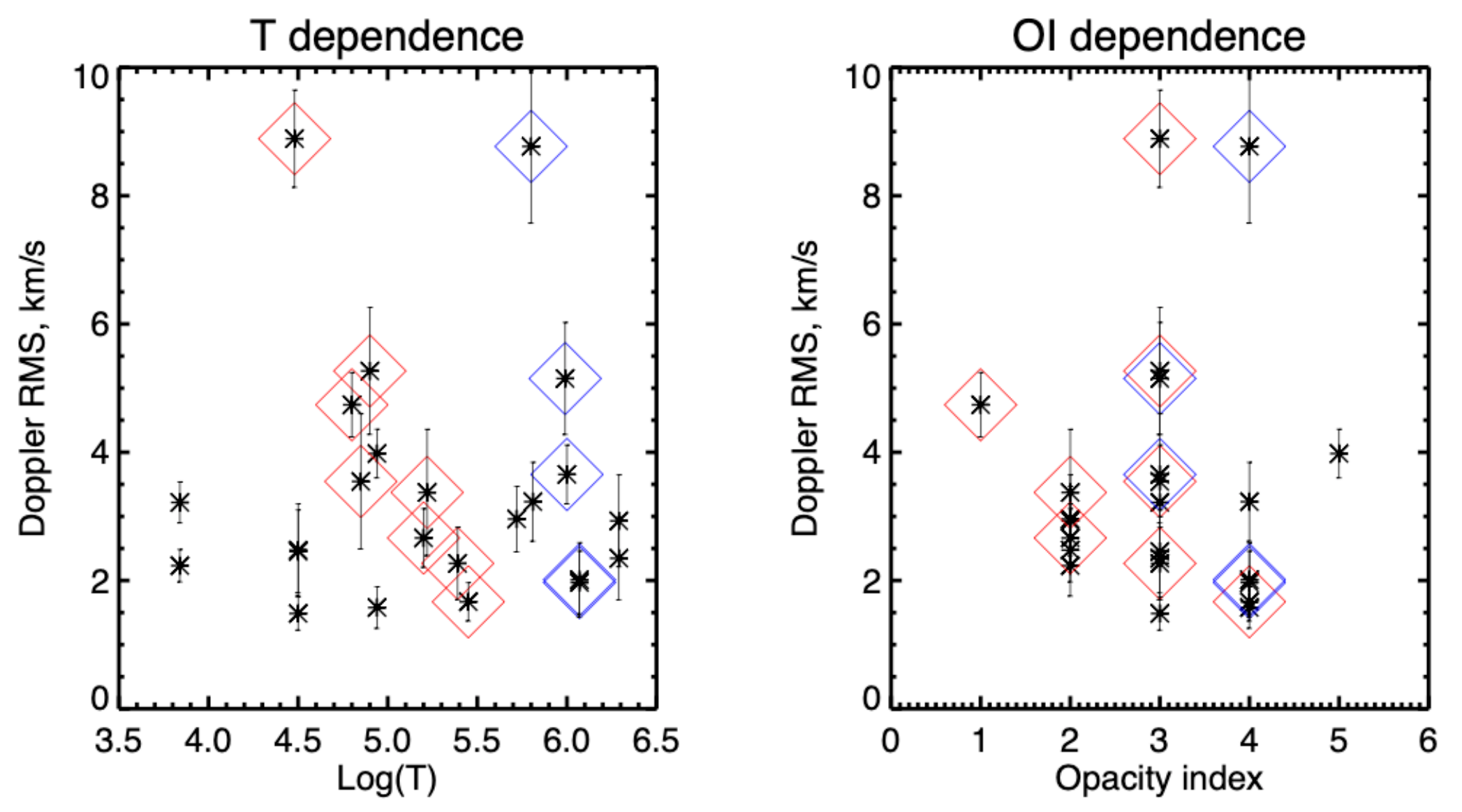}
\caption{\textit{RMS Doppler amplitudes in km/s for the all lines studied (0.1--50~MHz), plotted against the CHIANTI log(\textit{T})~[K] at G(T) maximum (left)  and our ``opacity index'' OI (right), as defined in Section~\ref{sec:blends}}.
The uncertainty ranges shown here and below are the sample standard deviations.
The colored diamond symbols mark the lines of the O series (red) and the Mg series (blue).
The outliers in the right panel are O~{\sc ii} 718~\AA\ and Mg~{\sc vii} 434~\AA.}
\label{fig:logt_oi}
\end{figure}

Figure~\ref{fig:logt_oi} shows the Doppler RMS values of all lines in our list, integrated above the 0.1~mHz limit (left panel), calling out the multiple ionization states for oxygen and magnesium.
The right panel shows the same values as a function of our opacity index (Table~\ref{tab:linelist}).
This shows discordant lines (O~{\sc ii} 718~\AA\ and Mg~{\sc vii} 434~\AA), likely the result of the crudeness of this index.

\subsection{The chromospheric lines}\label{sec:chrom}
The EVE linelist includes three Lyman-series lines ($\beta, \gamma, \delta$) also available in the EVE/MEGS-B database, as well as two He~{\sc i} lines.
These uniquely (in our linelist) have chromospheric contributions.

\begin{figure}
\centering    
   \includegraphics[width=\linewidth]{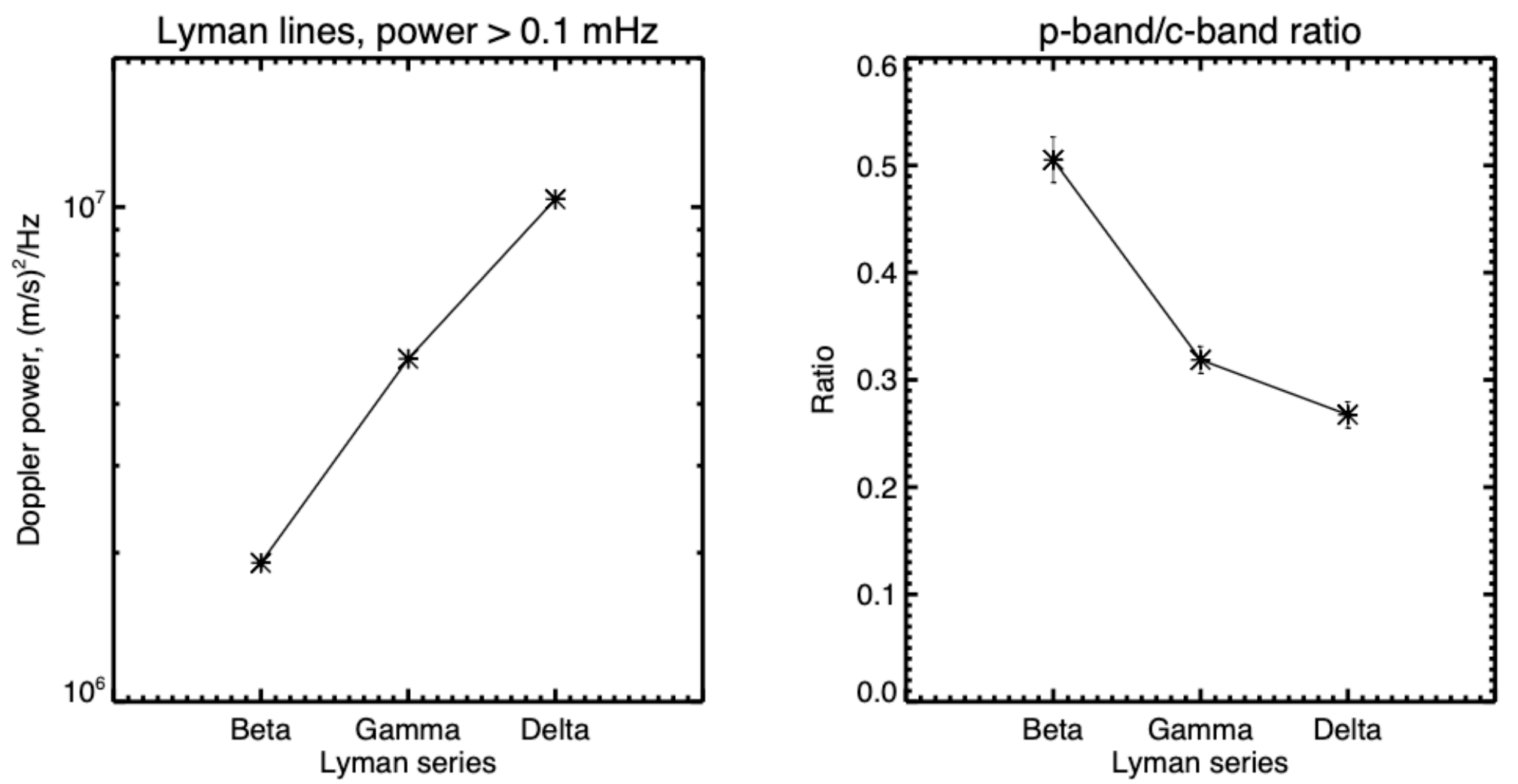}
\caption{\textit{Left, Doppler powers ($>0.1$~mHz) for the higher Lyman lines of hydrogen; right, the ratio of p-band (1-5~mHz) to c-band (5-20~mHz) power levels.}
}
\label{fig:hlines_plot}
\end{figure}

Figure~\ref{fig:hlines_plot} shows the $>0.1$~mHz Doppler amplitudes for the three Lyman lines (left panel) and the ratio of p-band (1-5~mHz) to c-band power (5-20~mHz) as in Figure~\ref{fig:pbandcband}.

\begin{figure}
\centering    
   \includegraphics[width=0.49\linewidth]{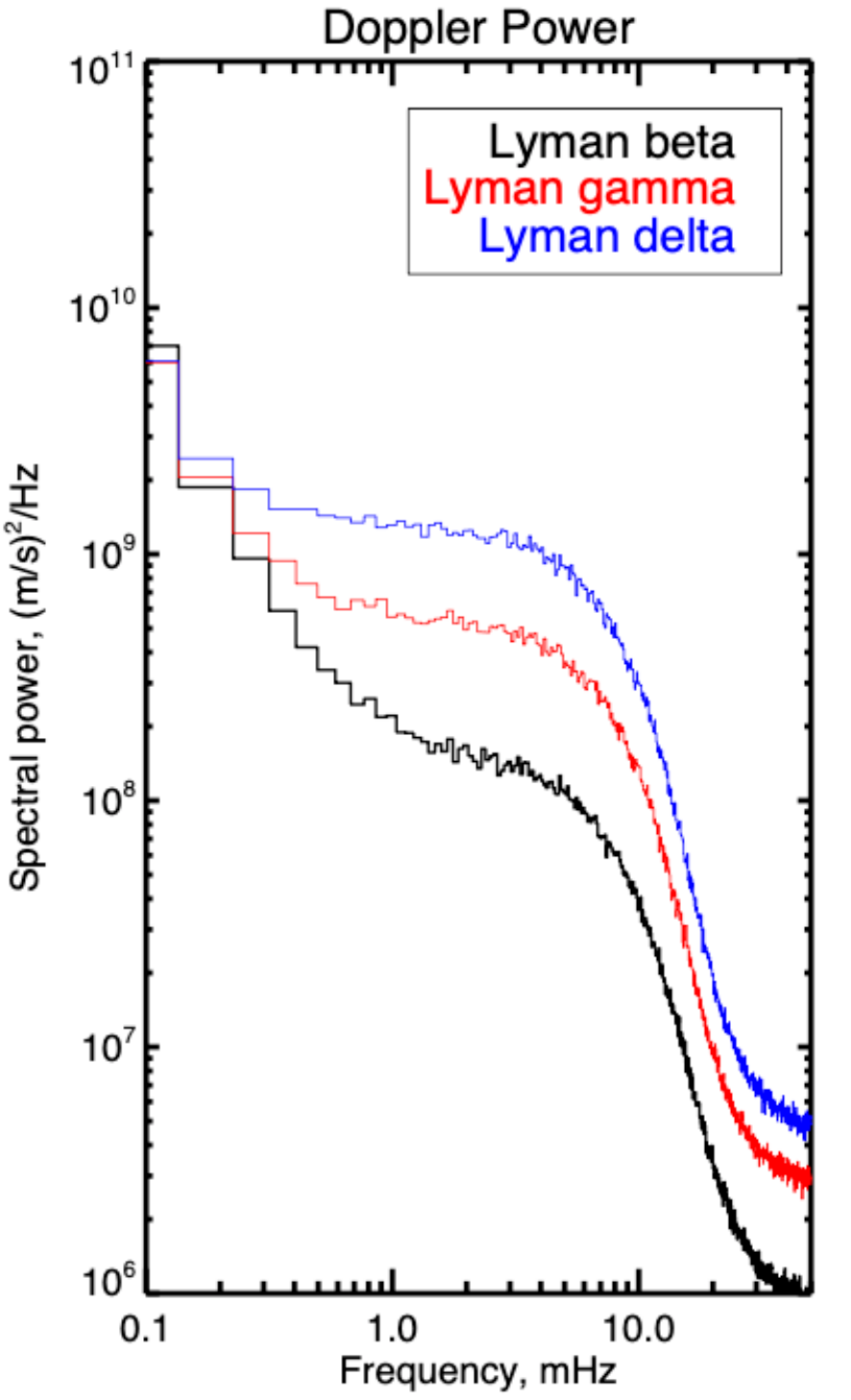}
   \includegraphics[width=0.49\linewidth]{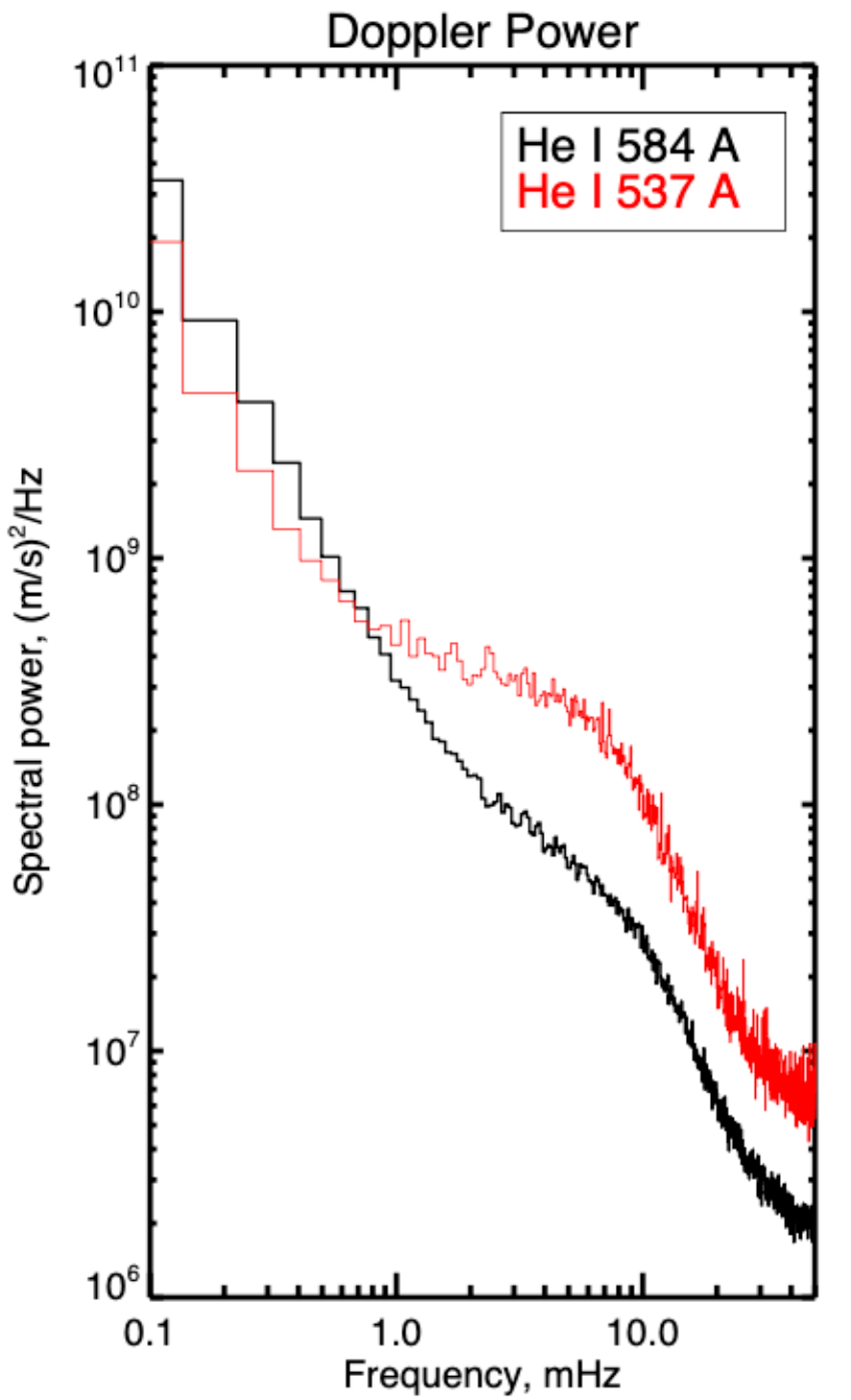}
\caption{\textit{Comparisons of spectral power for the three Lyman lines and the two He~{\sc i} lines in our list.
These lines have our crude opacity index (Table~\ref{tab:linelist}) OI values [3,3,2] and [3,2], respectively, thus corroborating the suppression of the 5~mHz spectral power for the high-opacity lines.}
}
\label{fig:hhe_lines}
\end{figure}

We compare Doppler spectra of the available H and He lines in Figure~\ref{fig:hhe_lines}.
These show a striking increase in the 5~mHz Harvey spectrum, when we compare Lyman~$\delta$ with Lyman~$\beta$, and likewise greater 5~mHz power for the 537~\AA\ line of He~I line as compared with 584~\AA.
These properties presumably result from the large optical depths in these lines.

\subsection{A First Ionization Potential (FIP) search}\label{sec:FIP}

\begin{figure}
\centering    
   \includegraphics[width=\linewidth]{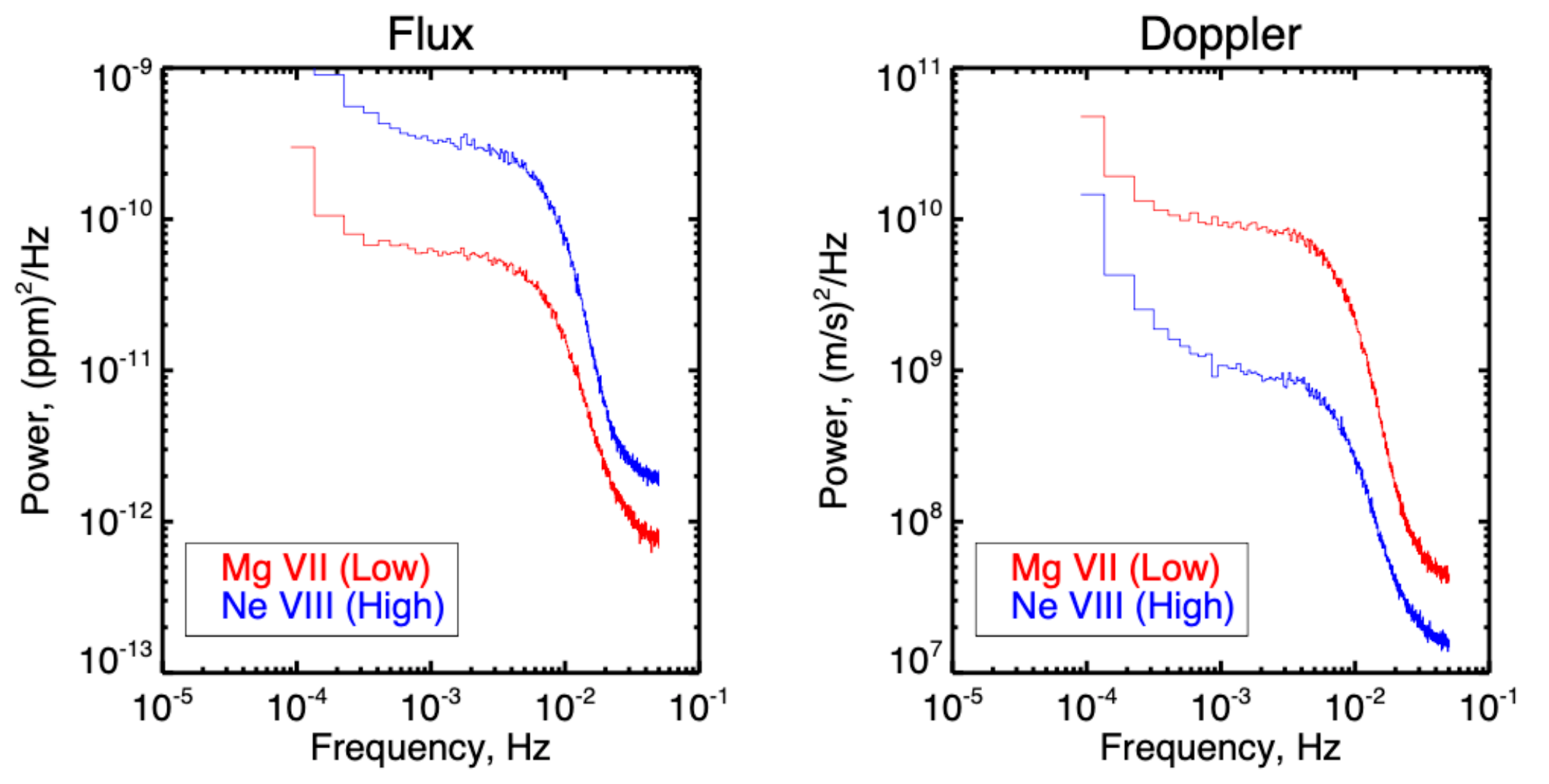}
\caption{\textit{Power spectra for a FIP comparison line pair, with high-FIP (Ne~{\sc viii}) in blue and low-FIP (Mg~{\sc vii}) in red.}
}
\label{fig:show_FIP}
\end{figure}

The line list (Table~\ref{tab:linelist}) contains one line pair suitable for checking whether elemental abundances can produce recognizeable signatures in the power spectra: Mg~{\sc vii} 434.92~\AA\ and Ne~{\sc viii} 770.409~\AA, with log~\textit{T}~[K] = 5.8 (0.63~MK) by CHIANTI.
To detect a FIP bias\footnote{The ``first ionization potential'' effect  can segregate regions with different compositions.
For Ne and Mg, first ionization energies are 21.56 eV and 7.65 eV, respectively.} in Sun-as-a-star spectroscopy, if possible, would have broad implications  \citep{2005A&A...439..361Y}.
These two lines do indeed have very different power spectra, as shown in Figure~\ref{fig:show_FIP}.
We see that the ratio of flux to Doppler power levels has an order-of-magnitude difference for these two lines. 
Also note that the spectra for the Mg line have a more significant excess below 1~mHz; this matches the relative weakness of granulation scales that we have previously noted in lines formed in the corona.

\begin{figure}
\centering    
   \includegraphics[width=0.7\linewidth]{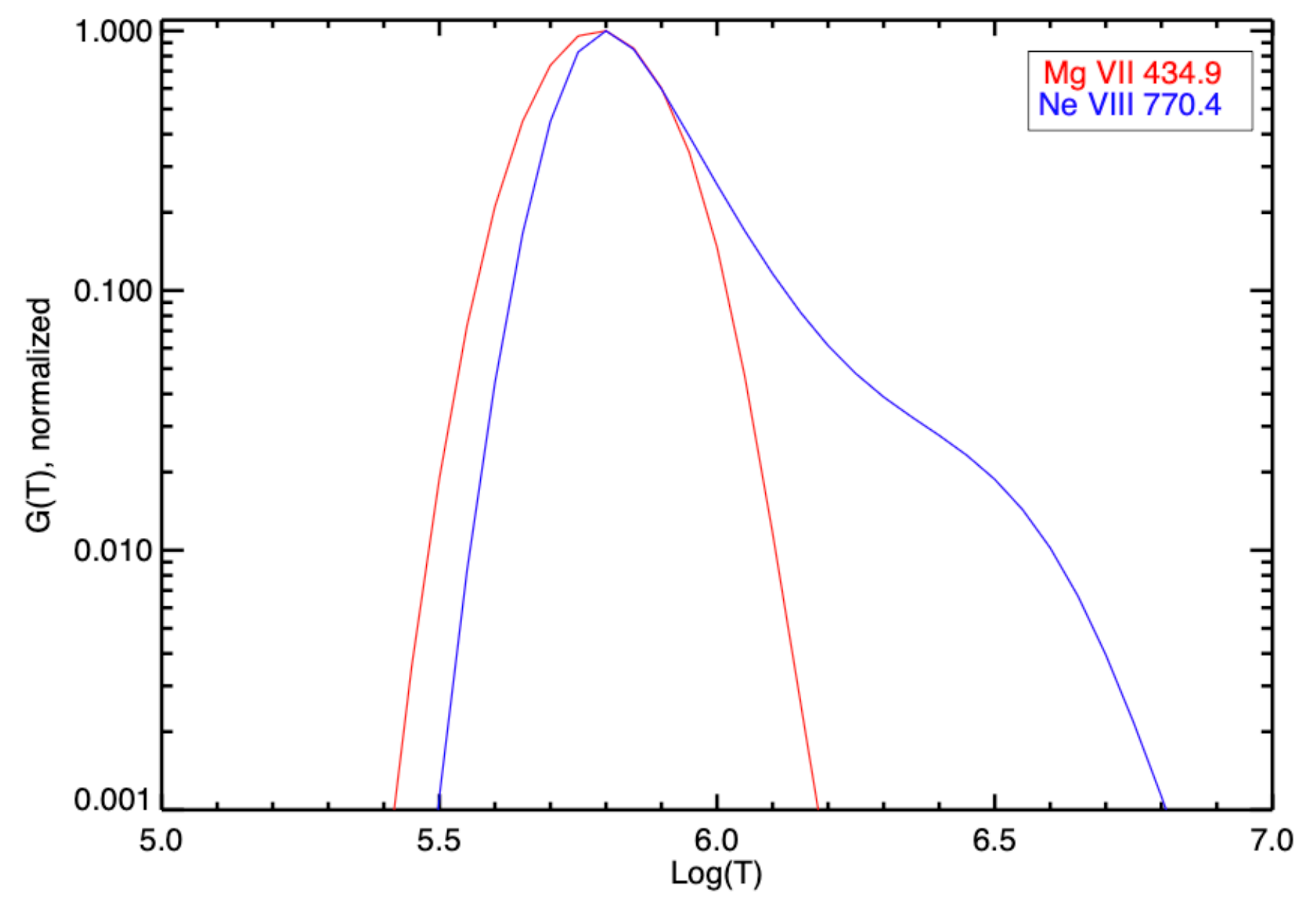}
\caption{\textit{The contribution functions G(T) for the FIP spectral line pair Mg~{\sc vii} and Ne~{\sc viii}, with the latter being lithium-like and therefore having broader contributions.
This line pair (see the right panel of Figure~\ref{fig:show_FIP}) also matches the OI estimations in Table~\ref{tab:linelist} regarding the suppression of the Harvey spectrum in the corona. 
Our linelist (Figure~\ref{fig:master_goft}) contains several lines in this sequence.}
}
\label{fig:FIP_GOFTs}
\end{figure}

A closer look at the full contribution functions G(T) \citep{1997A&AS..125..149D} for these two lines (Fig.~\ref{fig:FIP_GOFTs}) immediately suggests that the behavior of the Doppler spectra results from a high-temperature tail of the Ne contribution function.
This means that the Ne line is dominated by coronal velocity fields, rather than FIP bias, and is consistent with the idea that the coronal variability has a much weaker contribution from granulation-scale driving motions.
\cite{2005A&A...439..361Y} used imaging spectroscopy from the CDS/NIS instrument \citep{1995SoPh..162..233H} for multiple Ne and Mg lines to detect a weak fractionation effect across cell boundaries, but it appears that one cannot detect it in these Sun-as-a-star power spectra because of competition with the dynamical effects.
In addition, the EVE spectra do not have a useful spread of ionization states for these particular elements.

\section{Interpretation}

The EUV emission lines observed by EVE give us a comprehensive view of the Sun-as-a-star Doppler amplitude of plasma motions from the chromosphere to the low corona. 
The observations averaged over a single year (2011) have an excellent signal-to-noise ratio for the power spectrum in the 0.1-50~mHz range, which matches the behavior described by the \cite{1985ESASP.235..199H} function via its generalization.
The spectra generally do not show the $\langle v \rangle^2\, \propto\, f^{-5/3}$ slope expected from Kolmogorov equilibrium, although such a spectrum may exist underneath the Harvey-function contributions.
A quantitative theoretical understanding of the observed spectra will be complicated, involving contribution functions mixing optically-thick and optically-thin components.
Another particularly difficult observational problem has to do with solar active regions.
These contribute strongly to the emissions in some of the lines we have studied, but we do not have systematic monochromatic images in these lines that could help to define their distribution.

Figure~\ref{fig:logt_oi} (left panel) shows Doppler amplitudes of all lines in our list, integrated above the 0.1~mHz limit, calling out the multiple ionization states for oxygen and magnesium.
This plot compares mean line-of-sight velocities with the CHIANTI values of log~\textit{T}~[K] with uncertainties 
derived empirically from the statistics of the 3-hour chunk spectra across 2011.
The temperature dependence found in our line sequences could be  interpreted as a height dependence between the upper chromosphere and the transition region, but, in the absence of global Sun-as-a-star modeling, we cannot make this result very quantitative.
The results for the chromospheric lines (H and He) corroborate this pattern.
Based on multi-year time series from GOES as well as EVE, \cite{2020SpWea..1802331M} find that the quiet Sun (as opposed to active regions) dominates the solar Ly$\alpha$ emission, and so we expect that the higher members of the Lyman series behave similarly.

The absence of an obvious p-mode signature is informative, in light of the presence of this signal in the coronal power spectra obtained from CoMP observations reported in \cite{2007Sci...317.1192T}, and from DKIST/CRYO-NIRSP observations reported \cite{2025ApJ...986L...6M}. 
Both of these studies use data in the 1074.7~nm line of Fe~{\sc xiii} (log~\textit{T}~[K]) = 6.2) and refer to small fields of view.

\cite{2021ApJ...916...66B} suggest that the supergranulation signature in the intensity variability spectrum below about 1~mHz may arise not from convective motions, but from bright points within the network \cite[e.g.][]{2012ApJ...752...48C}.
Our observations show a similar pattern in the Doppler spectra, which might complicate this picture.

\section{Conclusions}\label{sec:concl}

The behavior of the power-spectral continuum variability of the Sun in the mHz band has received little study, perhaps because the p-mode resonances themselves are so interesting for both solar and stellar physics.
The EVE data we have described provide an excellent Sun-as-a-star database for characterizing the behavior of this continuum.
In this article we have only introduced the spectra and have not been able to compare the observations with detailed modeling and/or theoretical interpretation.
The 3-hour data chunks that we have analyzed allowed us to extend the coverage of the convective continuum to relatively high frequencies, generally not reaching the expected white-noise floor due to sensitivity limits, and averaging the power spectra over the entire year 2011 gives spectra with good SNR.

The EUV power spectra strongly resemble those of photospheric convective motions, showing broad-band continua rather than p-modes or chromospheric 3-minute oscillations.
We suggest that this morphological difference results from Sun-as-a-star emphasis on horizontal motions well observed near the limb.
The ready detectability of these spectra suggests that similar long-term stellar observations, especially longward of 91.1~nm, could usefully characterize stellar convection.

We draw several specific conclusions regarding the Doppler analysis of the EVE/MEGS-B emission lines:
\begin{enumerate}
\item The data provide useful power spectra both in intensity and in radial velocity out to the 50~mHz Nyquist-frequency limit of the sampling. 
We observe true solar variability at our highest frequencies, with a relatively low white-noise floor.
\item All MEGS-B spectral lines show the presence of a broad Doppler continuum following the generalized Harvey function suggested by photospheric convective motions, with a rollover frequency near 7~mHz.
\item The substantial power observed below the nominal acoustic cutoff frequency \citep{1973ApJ...184L.131C} confirms the efficiency of ``portals'' \cite{2007Sci...317.1192T} for the entry of low-frequency acoustic power into the transition region.
\item There is no spectral peak in the chromospheric 3~mHz range.
\item Lines with appreciable opacity in the corona show significantly reduced high-frequency (above 1~mHz) Doppler contributions.
\item The available series of O and Mg ionization states each show a significant range of amplitudes, with decreasing Harvey-function contributions at higher temperatures but reasonable agreement at the 0.1~mHz lower frequency limit of the analysis.
\item The chromospheric lines (H and He) also show diminishing Harvey-spectrum contributions at higher opacities.
\item The EVE Doppler spectra set an upper limit of about $10^{-5}$ for the ratio of Kolmogorov turbulent energy density to total kinetic energy density in the corona.
\end{enumerate}

The EVE data generally have good precision at most wavelengths, but quantitative interpretation is difficult in the absence of a global modeling framework.
The generation of Sun-as-a-star model power spectra is a prerequisite for the line-to-line differences that we see.
The power spectra are very precise and global in nature.
They require long integrations, but we believe that much can be learned about the mean solar atmosphere by detailed comparisons among lines and along the time series.
We anticipate further observational studies of aspects of EVE database
not considered in this article:
\begin{itemize}
    \item Study solar-cycle properties and rotational modulation (active regions).
    \item Explore the MEGS-A lines, generally at higher formation temperatures.
    \item Search for p-mode signatures (resonances) with the longer contiguous data sequences in the EVE database.
    \item Make detailed fits to Harvey-type functions and compare with quantitative models.
    \item Obtain a more rigorous description of coronal plasma turbulence.
    \item Characterize the supergranulation spectra below 100~$\mu$Hz.
    \item Explore the Doppler variability on active-region time scales below 1~$\mu$Hz.
\end{itemize}

\bigskip
\section{ACKNOWLEDGMENTS}

We are particularly grateful to Phil Scherrer and Tom Woods, who advised us on the intricacies of their respective datasets (HMI and EVE, respectively). 
We have benefited greatly from the referee's report on the original submission of this article.
Author Hudson thanks the University of Glasgow for hospitality during this work. 
L. Fletcher and S. Mulay acknowledge support from UK Research and Innovation’s Science and Technology Facilities Council under grant award number ST/X000990/1. 
A.-M. Broomhall acknowledges support from UK Research and Innovation’s Science and Technology Facilities Council under grant award number ST/X000915/1.

\bigskip
\section{DATA AVAILABILITY}

This article uses data from the EVE Level-2 database, accessible at \url{https://lasp.colorado.edu/eve/data_access/index.html}, interpreted via the CHIANTI atomic-physics database \citep{1997A&AS..125..149D,2019ApJS..241...22D} available at \url{https://www.chiantidatabase.org.}
Our SDO ephemeris data come from the very convenient JPL Horizons service (\url{https://ssd.jpl.nasa.gov/horizons}).

\newpage
\bibliographystyle{mnras}
\bibliography{EUV_mHz}

\newpage
..
\newpage
\noindent APPENDIX

Table~\ref{tab:app1} summarizes the Doppler timeseries behavior of the 26 chosen lines for EVE's 2011 data in broad spectral bands: the low frequencies (0.1-1~mHz) partially capturing supergranulation scales, a ``p-band'' (1-5~mHz) containing possible p-mode power and convection at granulation scales, a ``c-band'' (5-20~mHz) containing an uncharacterized convective continuum, a ``t-band'' containing the 20-45~mHz tail of the Harvey spectrum, and a possibly diagnostic ``n-band'' (45-50~mHz) sensitive to any high-frequency noise level, but also containing the highest available frequencies of true solar variability. 
The Tables lists Doppler power sum variances in (m/s)$^2$ for each line, for the entire frequency range and by spectral band.

In the Table, the uncertainty estimates show the RMS scatter of the power levels for individual bands, based on the number of chunk spectra available (typically 150-200).

\newpage\pagestyle{empty}
\begin{landscape}
\begin{table}
\caption{Doppler power totals for 2011 EVE MEGS-B data}
\label{tab:app1}
\begin{tabular}{l r l l l l l l}
\hline
  Ion & $\lambda$ (\AA) & 0.1-50 mHz& 0.1-1 mHz& 1-5 mHz& 5-20 mHz & 20-45 mHz& 45-50 mHz \\
      &  & (m/s)$^2$ & (m/s)$^2$ & (m/s)$^2$ & (m/s)$^2$ & (m/s)$^2$ & (m/s)$^2$ \\
\hline
  C III &    977.156 & 2.49e+06$\pm$1.1e+05 & 3.86e+05$\pm$1.8e+04 & 4.28e+05$\pm$6.9e+03 & 3.84e+05$\pm$5.6e+03 & 4.56e+04$\pm$1.7e+03 & 6.00e+03$\pm$2.4e+02 \\
 Fe XVI &    335.410 & 1.52e+08$\pm$7.7e+06 & 1.62e+07$\pm$8.7e+05 & 6.95e+07$\pm$3.5e+06 & 5.80e+07$\pm$3.1e+06 & 1.97e+06$\pm$1.3e+05 & 2.34e+05$\pm$1.7e+04 \\
  Fe XV &     69.682 & 6.48e+09$\pm$2.3e+08 & 1.12e+08$\pm$5.8e+06 & 5.22e+08$\pm$2.3e+07 & 1.96e+09$\pm$7.7e+07 & 3.20e+09$\pm$1.1e+08 & 6.26e+08$\pm$2.0e+07 \\
    H I &   1025.720 & 2.21e+06$\pm$6.7e+04 & 3.21e+05$\pm$1.2e+04 & 5.56e+05$\pm$7.1e+03 & 4.80e+05$\pm$5.1e+03 & 3.66e+04$\pm$1.1e+03 & 4.77e+03$\pm$1.6e+02 \\
    H I &    949.700 & 1.04e+07$\pm$1.0e+05 & 1.18e+06$\pm$2.8e+04 & 4.43e+06$\pm$5.0e+04 & 3.72e+06$\pm$3.5e+04 & 1.89e+05$\pm$2.9e+03 & 2.40e+04$\pm$4.4e+02 \\
    H I &    972.537 & 4.98e+06$\pm$6.3e+04 & 5.99e+05$\pm$1.5e+04 & 1.89e+06$\pm$2.1e+04 & 1.63e+06$\pm$1.4e+04 & 1.07e+05$\pm$1.6e+03 & 1.48e+04$\pm$2.9e+02 \\
   He I &    537.030 & 6.13e+06$\pm$5.2e+05 & 7.25e+05$\pm$5.7e+04 & 1.31e+06$\pm$9.9e+04 & 1.37e+06$\pm$1.3e+05 & 2.44e+05$\pm$4.4e+04 & 3.23e+04$\pm$5.7e+03 \\
   He I &    584.335 & 6.02e+06$\pm$4.2e+05 & 1.07e+06$\pm$6.9e+04 & 4.49e+05$\pm$2.1e+04 & 3.49e+05$\pm$1.3e+04 & 7.99e+04$\pm$5.7e+03 & 1.06e+04$\pm$8.4e+02 \\
  Mg IX &    368.071 & 1.33e+07$\pm$2.1e+05 & 1.58e+06$\pm$3.8e+04 & 5.52e+06$\pm$8.5e+04 & 4.58e+06$\pm$5.7e+04 & 2.69e+05$\pm$5.4e+03 & 3.43e+04$\pm$7.9e+02 \\
  Mg IX &    439.180 & 1.00e+09$\pm$2.1e+07 & 7.00e+07$\pm$1.8e+06 & 2.99e+08$\pm$5.5e+06 & 4.24e+08$\pm$1.0e+07 & 1.59e+08$\pm$5.1e+06 & 2.22e+07$\pm$5.7e+05 \\
Mg VIII &    436.730 & 2.65e+07$\pm$7.6e+05 & 2.49e+06$\pm$1.1e+05 & 8.78e+06$\pm$2.3e+05 & 9.75e+06$\pm$2.7e+05 & 2.06e+06$\pm$1.4e+05 & 2.75e+05$\pm$1.9e+04 \\
 Mg VII &    434.920 & 7.69e+07$\pm$1.4e+06 & 8.49e+06$\pm$2.3e+05 & 3.23e+07$\pm$6.2e+05 & 2.79e+07$\pm$4.9e+05 & 1.65e+06$\pm$5.5e+04 & 2.00e+05$\pm$7.3e+03 \\
   Mg X &    609.800 & 3.88e+06$\pm$2.4e+05 & 6.19e+05$\pm$3.7e+04 & 4.44e+05$\pm$1.4e+04 & 3.96e+05$\pm$1.4e+04 & 6.81e+04$\pm$5.8e+03 & 8.99e+03$\pm$7.8e+02 \\
   Mg X &    624.943 & 4.05e+06$\pm$3.3e+05 & 4.98e+05$\pm$4.0e+04 & 7.79e+05$\pm$4.1e+04 & 7.56e+05$\pm$4.0e+04 & 1.26e+05$\pm$1.2e+04 & 1.68e+04$\pm$1.4e+03 \\
  N III &    991.511 & 1.58e+07$\pm$1.4e+05 & 1.84e+06$\pm$4.0e+04 & 6.55e+06$\pm$8.1e+04 & 5.87e+06$\pm$5.7e+04 & 3.67e+05$\pm$5.4e+03 & 4.57e+04$\pm$8.3e+02 \\
Ne VIII &    770.409 & 1.04e+07$\pm$3.8e+05 & 1.18e+06$\pm$5.4e+04 & 3.44e+06$\pm$1.1e+05 & 3.32e+06$\pm$1.0e+05 & 5.27e+05$\pm$1.8e+04 & 7.63e+04$\pm$2.9e+03 \\
 Ne VII &    465.220 & 8.75e+06$\pm$2.6e+05 & 1.19e+06$\pm$4.4e+04 & 2.56e+06$\pm$4.0e+04 & 2.22e+06$\pm$3.1e+04 & 1.41e+05$\pm$4.7e+03 & 1.76e+04$\pm$6.5e+02 \\
  O III &    525.794 & 2.77e+07$\pm$9.8e+05 & 2.79e+06$\pm$1.1e+05 & 7.62e+06$\pm$2.5e+05 & 7.69e+06$\pm$2.4e+05 & 1.40e+06$\pm$3.5e+04 & 1.90e+05$\pm$5.3e+03 \\
  O III &    599.590 & 1.26e+07$\pm$1.1e+06 & 1.40e+06$\pm$1.2e+05 & 2.17e+06$\pm$1.5e+05 & 2.19e+06$\pm$1.5e+05 & 4.12e+05$\pm$3.4e+04 & 5.60e+04$\pm$4.8e+03 \\
  O III &    835.289 & 2.25e+07$\pm$2.5e+05 & 2.56e+06$\pm$5.6e+04 & 9.42e+06$\pm$1.3e+05 & 8.21e+06$\pm$9.7e+04 & 7.28e+05$\pm$1.6e+04 & 1.07e+05$\pm$2.9e+03 \\
   O II &    718.535 & 7.90e+07$\pm$5.8e+05 & 8.23e+06$\pm$1.7e+05 & 3.32e+07$\pm$3.3e+05 & 2.95e+07$\pm$2.2e+05 & 3.54e+06$\pm$3.6e+04 & 5.54e+05$\pm$6.9e+03 \\
   O IV &    554.513 & 1.14e+07$\pm$9.7e+05 & 1.18e+06$\pm$1.1e+05 & 1.62e+06$\pm$1.1e+05 & 1.87e+06$\pm$1.1e+05 & 5.26e+05$\pm$3.0e+04 & 7.26e+04$\pm$4.5e+03 \\
   O IV &    790.199 & 7.09e+06$\pm$2.1e+05 & 8.61e+05$\pm$3.0e+04 & 2.30e+06$\pm$6.8e+04 & 2.14e+06$\pm$6.1e+04 & 2.78e+05$\pm$1.1e+04 & 3.96e+04$\pm$1.9e+03 \\
   O VI &   1031.914 & 2.78e+06$\pm$8.9e+04 & 4.26e+05$\pm$1.8e+04 & 6.55e+05$\pm$9.4e+03 & 5.54e+05$\pm$6.6e+03 & 3.21e+04$\pm$1.1e+03 & 3.93e+03$\pm$1.6e+02 \\
    O V &    629.732 & 5.14e+06$\pm$3.2e+05 & 8.64e+05$\pm$4.9e+04 & 5.42e+05$\pm$1.7e+04 & 5.08e+05$\pm$1.5e+04 & 1.17e+05$\pm$6.7e+03 & 1.73e+04$\pm$1.1e+03 \\
 Si XII &    499.406 & 5.51e+06$\pm$4.2e+05 & 6.94e+05$\pm$4.9e+04 & 1.06e+06$\pm$5.5e+04 & 1.07e+06$\pm$6.1e+04 & 1.76e+05$\pm$2.0e+04 & 2.48e+04$\pm$3.0e+03 \\
 Si XII &    521.000 & 8.58e+06$\pm$5.1e+05 & 9.73e+05$\pm$6.3e+04 & 1.95e+06$\pm$9.8e+04 & 2.15e+06$\pm$1.3e+05 & 3.96e+05$\pm$4.5e+04 & 5.37e+04$\pm$6.6e+03 \\

  \end{tabular}
\end{table}
\end{landscape}

\end{document}